\documentclass[manuscript,screen]{acmart}
\usepackage[ruled,linesnumbered]{algorithm2e}
\usepackage{graphicx}
\usepackage{array}
\usepackage[caption=false,font=normalsize,labelfont=sf,textfont=sf]{subfig}
\usepackage{textcomp}
\usepackage{stfloats}
\usepackage{url}
\usepackage{verbatim}
\usepackage{graphicx}
\usepackage{enumitem}
\usepackage{multirow}
\usepackage{colortbl}
\usepackage{booktabs}
\usepackage{makecell}
\usepackage{threeparttable}
\usepackage{calc}
\usepackage{colortbl}
\usepackage{color}
\usepackage{url}
\usepackage{balance}
\usepackage{breakurl}
\usepackage{listings}
\usepackage{hyperref}
\usepackage{fontawesome}
\usepackage[most]{tcolorbox}

\definecolor{codegreen}{RGB}{199, 237, 204}
\definecolor{codered}{RGB}{253, 230, 224}
\definecolor{codeblue}{RGB}{220, 226, 241}
\lstdefinelanguage{diff}{
frame=shadowbox,
breaklines,
basicstyle=\small,
escapeinside={\%*}{*)},
alsoletter={.},
keepspaces=true,
captionpos=b,
columns=fullflexible,
numberstyle=\footnotesize,
numbersep=5pt,
breaklines=true,
abovecaptionskip=1em,
belowcaptionskip=0em
}
\AtBeginDocument{%
  }

\setcopyright{acmlicensed}
\copyrightyear{2018}
\acmYear{2018}
\acmDOI{XXXXXXX.XXXXXXX}

\acmConference[Conference acronym 'XX]{Make sure to enter the correct
  conference title from your rights confirmation emai}{June 03--05,
  2018}{Woodstock, NY}
\acmISBN{978-1-4503-XXXX-X/18/06}

\begin{document}

\title{An Empirical Study on the Potential of LLMs in Automated Software Refactoring}

\author{Bo Liu}
\affiliation{%
  \institution{Beijing Institute of Technology}
  \city{Beijing}
  \country{China}
}
\email{liubo@bit.edu.cn}

\author{Yanjie Jiang}
\affiliation{%
  \institution{Peking University}
  \city{Beijing}
  \country{China}
}
\email{yanjiejiang@pku.edu.cn}

\author{Yuxia Zhang}
\affiliation{%
  \institution{Beijing Institute of Technology}
  \city{Beijing}
  \country{China}
}
\email{yuxiazh@bit.edu.cn}

\author{Nan Niu}
\affiliation{%
  \institution{University of Cincinnati}
  \city{Cincinnati}
  \country{USA}
}
\email{nan.niu@uc.edu}

\author{Guangjie Li}
\affiliation{%
  \institution{National Innovation Institute of Defense Technology}
  \city{Beijing}
  \country{China}}
\email{liguangjie_er@126.com}

\author{Hui Liu}
\authornote{Corresponding author}
\affiliation{%
  \institution{Beijing Institute of Technology}
  \city{Beijing}
  \country{China}
}
\email{liuhui08@bit.edu.cn}

\renewcommand{\shortauthors}{Liu et al.}

\begin{abstract}
  Software refactoring is an essential activity for improving the readability, maintainability, and reusability of software projects. To this end, a large number of automated or semi-automated approaches/tools have been proposed to locate poorly designed code, recommend refactoring solutions, and conduct specified refactorings. However, even equipped with such tools, it remains challenging for developers to decide where and what kind of refactorings should be applied. Recent advances in deep learning techniques, especially in large language models (LLMs), make it potentially feasible to automatically refactor source code with LLMs. However, it remains unclear how well LLMs perform compared to human experts in conducting refactorings automatically and accurately. To fill this gap, in this paper, we conduct an empirical study to investigate the potential of LLMs in automated software refactoring, focusing on the identification of refactoring opportunities and the recommendation of refactoring solutions. We first construct a high-quality refactoring dataset comprising 180 real-world refactorings from 20 projects, and conduct the empirical study on the dataset. With the to-be-refactored Java documents as input, ChatGPT and Gemini identified only 28 and 7 respectively out of the 180 refactoring opportunities. However, explaining the expected refactoring subcategories and narrowing the search space in the prompts substantially increased the success rate of ChatGPT from 15.6\% to 86.7\%. Concerning the recommendation of refactoring solutions, ChatGPT recommended 176 refactoring solutions for the 180 refactorings, and 63.6\% of the recommended solutions were comparable to (even better than) those constructed by human experts. However, 13 out of the 176 solutions suggested by ChatGPT and 9 out of the 137 solutions suggested by Gemini were unsafe in that they either changed the functionality of the source code or introduced syntax errors, which indicate the risk of LLM-based refactoring. To this end, we propose a detect-and-reapply tactic, called \texttt{RefactoringMirror}, to avoid such unsafe refactorings. By reapplying the identified refactorings to the original code using thoroughly tested refactoring engines, we can effectively mitigate the risks associated with LLM-based automated refactoring while still leveraging LLM's intelligence to obtain valuable refactoring recommendations. Our evaluation results suggest that \texttt{RefactoringMirror} accurately identified and reapplied 94.3\% of the refactorings conducted by LLMs, and successfully avoided all of the buggy solutions.
\end{abstract}

\ccsdesc[500]{Software and its engineering}

\keywords{Software Refactoring, Large Language Model, Empirical Study, Software Quality}


\maketitle

\section{Introduction}
\label{section:introduction}
Software refactoring is widely employed to improve software quality, especially its readability, maintainability, and reusability~\cite{fowler1999refactoring,mens2004survey,baqais2020automatic}. To facilitate software refactoring, a large number of approaches/tools have been proposed to identify refactoring opportunities~\cite{tourwe2003identifying,tsantalis2009identification,tsantalis2011identification},
to recommend refactoring solutions~\cite{mkaouer2017robust,alizadeh2018reducing,alizadeh2018interactive}, and to automatically conduct specified refactorings~\cite{ge2012reconciling,foster2012witchdoctor,alizadeh2019refbot}. Refactoring engines like IntelliJ IDEA~\cite{idea2024refactoring} and JDeodorant~\cite{fokaefs2007jdeodorant,tsantalis2008jdeodorant} have been successfully adopted to automate the execution of refactorings. However, before the refactorings could be executed automatically, developers should explicitly specify which part of the source code should be refactored, what kind of refactorings should be applied, and the detailed parameters of the refactorings (e.g., a new method name for \emph{rename method} refactoring). Even with the support of the state-of-the-art refactoring approaches/tools, it remains challenging and time-consuming for developers to make such decisions~\cite{feitelson2020developers,peruma2022refactor}, which in turn prevents software refactoring from reaching its maximal potential.

Large language models (LLMs), like GPT-4~\cite{achiam2023gpt} and Gemini~\cite{team2023gemini}, have the ability to learn complex and massive knowledge~\cite{liu2023summary}. Consequently, LLMs emerge as a potential solution that could significantly advance automated software refactoring. It is reported that LLMs have demonstrated promising results in various software engineering tasks, like code generation~\cite{mu2023clarifygpt}, fault location~\cite{wu2023large}, and program repair~\cite{xia2023keep}. 
LLMs' impressive capability in understanding and generating natural languages and source code makes it potentially feasible to automatically refactor source code with LLMs. However, it remains unknown whether LLMs can conduct refactorings automatically, whether the refactorings suggested/conducted by LLMs are of high quality, and whether refactorings conducted by LLMs are reliable. 

To fill this gap, in this paper, we conduct a comprehensive empirical study to obtain a profound insight into the challenges and opportunities of LLM-based refactoring as well as to understand the extent to which LLMs can automate software refactoring to alleviate the burden on developers.  We select two representative LLMs, i.e., GPT-4 and Gemini, for the study, and answer the following two research questions:
\begin{itemize}
    \item RQ1: How well do LLMs (GPT-4 and Gemini) work in the identification of refactoring opportunities?
    \item RQ2: How well do GPT-4 and Gemini work in the recommendation of refactoring solutions?
\end{itemize}
To answer the above questions, we first construct a high-quality refactoring dataset comprising 180 real-world refactorings from 20 projects. After that, we request LLMs to refactor the entire Java documents that contain the discovered refactoring opportunities. We conduct a quantitative analysis of LLMs's effectiveness in identifying refactoring opportunities. We also request three human experts to manually and independently assess the quality of the refactoring solutions suggested by LLMs. 
Our evaluation results suggest that LLMs have the potential for automated software refactoring. However, their performance varies significantly among different types of refactorings, and thus we refine their performance by a taxonomy of refactorings. 
We also noticed that LLM-based refactoring could be risky. 22 out of the 313 solutions suggested by GPT-4 and Gemini were unsafe in that they either changed the functionality of the source code or introduced syntax errors. 

To improve the safety of LLM-based refactoring, we propose a detect-and-reapply tactic (called \texttt{RefactoringMirror}) to avoid unsafe refactorings conducted by LLMs. When a to-be-refactored source code (noted as $c$) is fed to LLMs, it generates an improved version of the code (noted as $c'$). \texttt{RefactoringMirror} takes $c$ and $c'$ as input, compares their differences, and identifies a list of refactorings that have been applied to $c$. It then reapplies the identified refactorings to $c$ by well-tested refactoring engines (like IntelliJ IDEA), and generates a reliable and improved version of the code (noted as $\hat{c}$). As a result, it not only keeps the beneficial refactorings but also substantially improves the safety of the refactorings, ensuring that no functional changes are introduced. Our evaluation results suggest that it successfully avoided all of the buggy refactoring solutions conducted by GPT-4 and Gemini.

In this paper, we make the following contributions:
\begin{itemize}
    \item \textbf{A comprehensive empirical study} on evaluating LLMs' potential in automated software refactoring. It reveals the strengths and weaknesses of LLMs in software refactoring, refined by a refactoring taxonomy. 
    \item \textbf{A detect-and-reapply tactic} to mitigate the risk of LLM-based software refactoring.
    \item  \textbf{A new high-quality refactoring dataset} validated with multiple tools and refactoring experts.
\end{itemize} 

The rest of the paper is structured as follows. Section~\ref{section:studydesign} details the study design. Section~\ref{section:RQ1} evaluates the capability of LLMs in identifying refactoring opportunities. Section~\ref{section:RQ2} evaluates the capability of LLMs in recommending refactoring solutions. Section~\ref{section:mitigation} otulines the overall workflow of \textsc{RefactoringMirror}. Section~\ref{section:discussion} delves into the threats to validity, limitations, and implications. Section~\ref{section:relatedwork} provides an overview of related work. Finally, Section~\ref{section:conclusion} concludes the paper and suggests directions for future work.

\section{Study Design}
\label{section:studydesign}
\subsection{Research Questions}
The empirical study is designed to evaluate the capabilities of large language models (LLMs) in automated software refactoring. Specifically, the study 
investigates the following research questions:
\begin{enumerate}[leftmargin=1.2cm]
    \item[\textbf{RQ1.}] How well do LLMs work in the automated identification of refactoring opportunities?
    \item[\textbf{RQ2.}] How well do LLMs work in the automated recommendation of refactoring solutions?
\end{enumerate}

In this study, we employ the latest GPT-4~\cite{gpt2024model} (called GPT for short in the rest of this paper) and Gemini-1.0 Pro~\cite{gemini2024model} (called Gemini for short) as the evaluated large language models because they represent the state of the art in this area~\cite{lu2024gpt,hu2024can}.
RQ1 concerns the capability of LLMs in refactoring opportunity identification. RQ1 can be further divided into three sub-research questions:
\begin{enumerate}[leftmargin=1.5cm]
    \item[\textbf{RQ1-1.}] How well do LLMs work in identifying refactoring opportunities without specifying concrete refactoring types?
    \item[\textbf{RQ1-2.}] How well do LLMs work in identifying refactoring opportunities when refactoring types are explicitly specified?
    \item[\textbf{RQ1-3.}] How does the size of the to-be-refactored source code influence the performance of LLMs in identifying refactoring opportunities?
    \item[\textbf{RQ1-4.}] In which cases do LLMs perform well or poorly in identifying refactoring opportunities? And how can prompt strategies be improved to identify more refactoring opportunities?
\end{enumerate}
By comparing RQ1-1 and RQ1-2, we can not only reveal the potential of LLMs, but also reveal to what extent explicitly specifying refactoring types can improve the performance of LLM-based refactorings. Notably, when the to-be-refactored source code is longer and more complex, it could be more challenging for LLMs to pick up the expected refactoring opportunities. To validate this hypothesis, we investigate RQ1-3 to reveal the correlation between the performance of LLM-based refactoring opportunity identification and the size of the to-be-refactored source code. 
By answering RQ1-4, we can gain a deeper understanding of the strengths and weaknesses of LLMs in identifying refactoring opportunities and whether the proposed prompt strategies can facilitate LLMs to identify more refactoring opportunities.

RQ2 concerns the capability of LLMs to suggest refactoring solutions. It can be further divided into three sub-research questions:
\begin{enumerate}[leftmargin=1.5cm]
    \item[\textbf{RQ2-1.}] How do the refactoring solutions suggested by LLMs compare to those conducted by human experts?
    \item[\textbf{RQ2-2.}] Are the suggested refactoring solutions safe? If not, how often are the solutions unsafe, and what is the problem with the suggested solutions?
\end{enumerate}
By answering RQ2-1, we may reveal the potential of LLMs in recommending refactoring solutions. Notably, as generic large language models do not validate the equivalence of the input (original source code) and output (source code after refactoring),  it is likely that the output could be semantically nonequivalent to the input, which results in changes in software system's external behaviors. However, according to the definition of software refactoring~\cite{fowler1999refactoring}, software refactoring should never change external behaviors. In this case, the suggested ``refactorings" are unsafe. Answering RQ2-2 would reveal how often the refactoring solutions suggested by LLMs are unsafe and why they are unsafe. 

\subsection{Dataset}
\label{section:dataset}
To conduct our study, we constructed a new high-quality refactoring dataset, called \emph{ref-Dataset}, due to the following reason: The existing datasets might have been utilized to train the selected LLMs. The new dataset was constructed to better evaluate the generalization capabilities of LLMs. To this end, we reused the 20 popular Java projects chosen by Grund et al.~\cite{grund2021codeshovel}. These projects contained rich evolution histories from the year 2000 to the present. They were from different domains (including code analysis, search engine, unit testing, etc.), which might help reduce the potential bias in the evaluation. Finally, they are publicly available, which facilitates the reproduction of our evaluation. However, project \texttt{junit4} did not have any code commits after April 2023, which indicates that all their data (including refactoring activities) may have been used as training data for GPT. Consequently, it was not used in our study. Notably, project \texttt{lucene-solr} has been split into two separate projects \texttt{lucene} and \texttt{solr}, and thus we used the two sub-projects in the empirical study.

From the selected projects, we discovered refactorings conducted on such projects.  
We applied the state-of-the-art refactoring detection tools (i.e.,  \texttt{ReExtractor}~\cite{liu2023automated,lyoubo2024reextractor} and \texttt{RefactoringMiner}~\cite{tsantalis2018accurate,tsantalis2020refactoringminer}) independently to mine refactorings starting from the latest commit of each project. These two tools were selected because 1) they represented the state of the art in refactoring detection; and 2) they supported the detection of both high-level refactorings (e.g., \emph{rename method}) and low-level refactorings (e.g., \emph{extract variable}) at commit level. Notably, we only mined refactorings from commits submitted after April 2023 because GPT's training data extends up to this date~\cite{gpt2024model}, while Gemini's training data extends to February 2023~\cite{gemini2024model}. Considering the capability of the employed refactoring detection tools and the popularity of refactorings~\cite{murphy2011we,negara2013comparative}, the empirical study focused on the following 8 within-document refactoring types: \emph{extract method}, \emph{extract variable}, \emph{inline method}, \emph{inline variable}, \emph{rename attribute}, \emph{rename method}, \emph{rename parameter}, and \emph{rename variable}. 
All such refactorings take a single Java document as input, and generate an improved version of it, which may simplify the prompt (and output) of LLMs.   
If either (or both) of the refactoring detection tools identified a refactoring of the selected refactoring types, we collected it as a candidate and requested three refactoring experts (noted as $PTC_A$) to validate each potential refactoring independently and manually. 
In case of inconsistent validation, we simply discarded the case and turned to the next one.
We collected the expert-validated refactorings into \emph{ref-Dataset}. If neither of the two tools identified a specific type of refactoring in a given project, we chronologically selected an expert-validated instance of the same refactoring type from other projects, and collected it into \emph{ref-Dataset}. As a result, \emph{ref-Dataset} comprises 180 real-world refactorings, encompassing 9 distinct types of refactorings from 20 open-source projects.

\begin{figure}[t]
    \centerline{\includegraphics[width=0.7\linewidth]{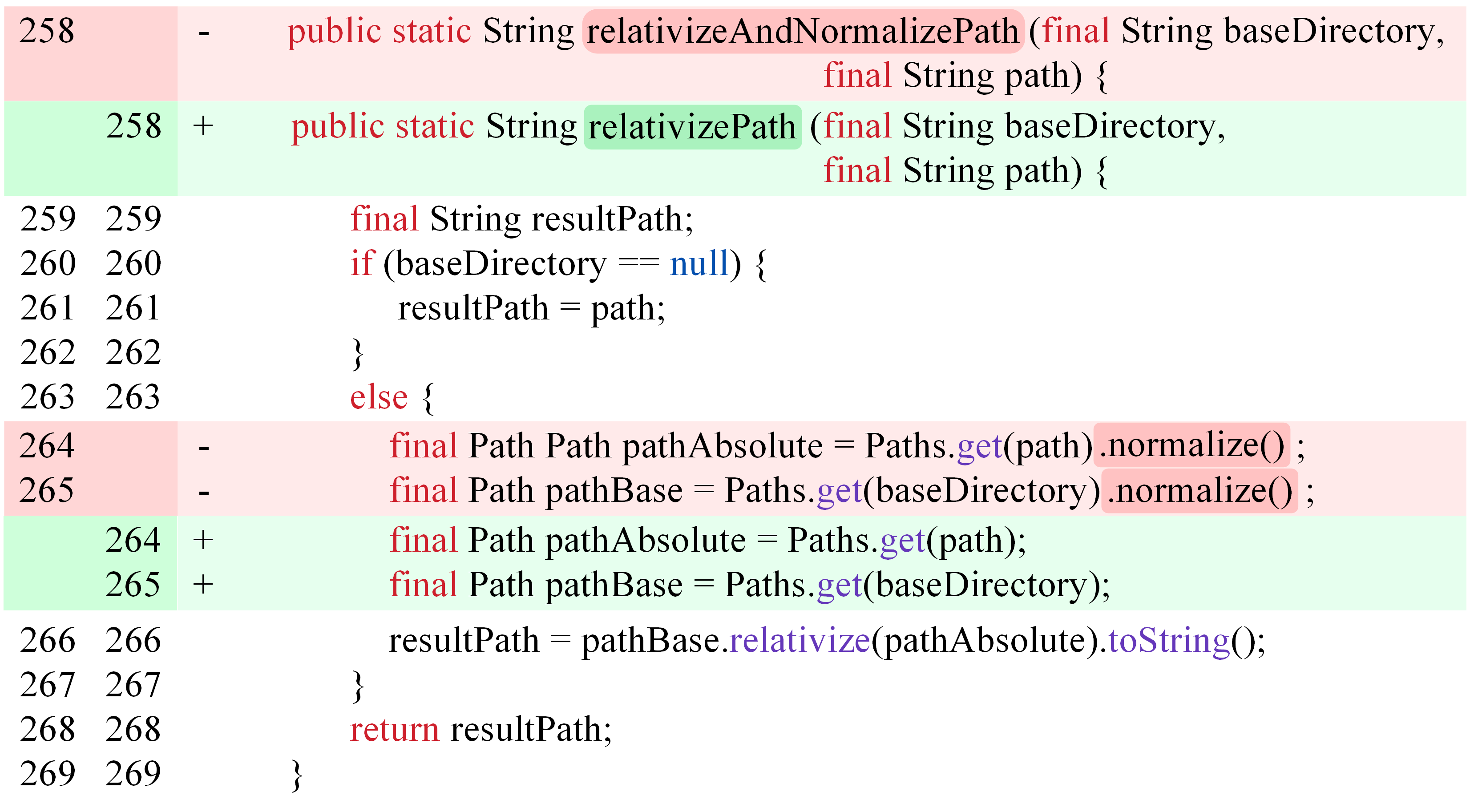}}
    \caption{An Example of Rename Method Refactorings from Commit d52eb5d in Checkstyle}
    \label{fig:example}
\end{figure}

\begin{figure}[t]
    \centerline{\includegraphics[width=0.6\linewidth]{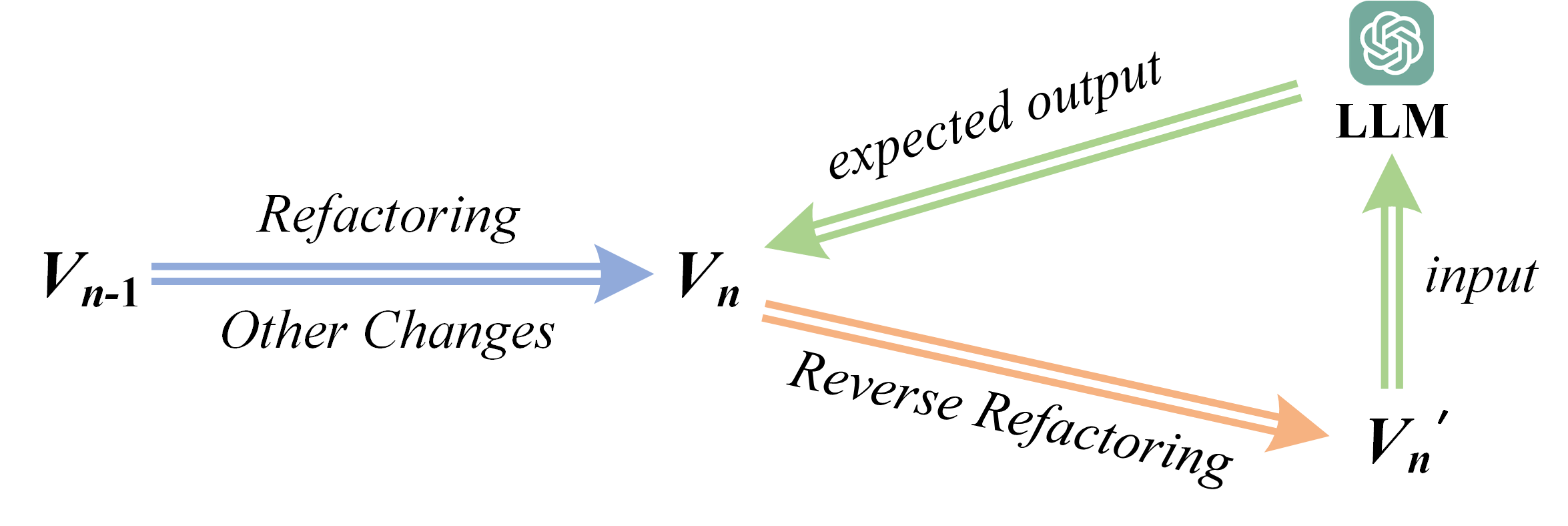}}
    \caption{Creating Input and Expected Output of LLMs}
    \label{fig:version}
\end{figure}

To validate the potential of LLM-based refactoring, we should feed source code containing refactoring opportunities and expect LLMs to generate the refactored version of the input. 
By discovering refactoring histories, we have $V_n$ and $V_{n-1}$ for each discovered refactoring in \emph{ref-Dataset} where $V_{n-1}$ represents the source code before refactoring and $V_{n}$  represents the source code after refactoring. An intuitive idea is to feed $V_{n-1}$ to LLMs, and expect them to generate $V_{n}$. However, it is impractical because of the following reasons. First, the commit changing $V_{n-1}$ to $V_{n}$ often contains other changes (e.g., functional changes and bug fixing) besides refactorings~\cite{silva2016we}. Consequently, LLMs with $V_{n-1}$ as input may not generate $V_{n}$ by applying refactorings only. Second, refactoring operations are often motivated by other non-refactoring changes. For example, 
\emph{rename method} refactoring is often conducted when developers are modifying the functionality of a method. A typical example is presented in Fig.~\ref{fig:example}. In this example, the developer removed the method call ``\emph{normalize}()" (Lines 264-265), and renamed the method from \emph{relativizeAndNormalizePath} to \emph{relativizePath} because the modified method body is inconsistent with the original method name.
Consequently, without such motivating non-refactoring changes (within the same commit), the refactorings may become unnecessary and thus LLMs cannot report such refactoring opportunities. 
To this end, we created suitable input (request) to LLMs with reverse refactoring as shown in Fig.~\ref{fig:version}. A reverse refactoring is a refactoring that removes an existing refactoring. For example, if a refactoring renames a method from \textit{oldName} to \textit{newName}, its corresponding reverse refactoring is to rename the same method from  \textit{newName} to \textit{oldName}. 
From the refactored version $V_{n}$, we manually applied reverse refactoring to remove the refactoring conducted by the given commit. As a result, ${V_n}'$ has all of the changes made by the commit except for the discovered refactoring. Consequently, it has all of the motivation changes (if there are any) as well as the refactoring opportunity. In conclusion, we take ${V_n}'$ as the input to the LLMs and ${V_n}$ as the expected output. 

\subsection{Prompt Template}
\label{section:setup}

\begin{figure}[t]
    \centerline{\includegraphics[width=0.7\linewidth]{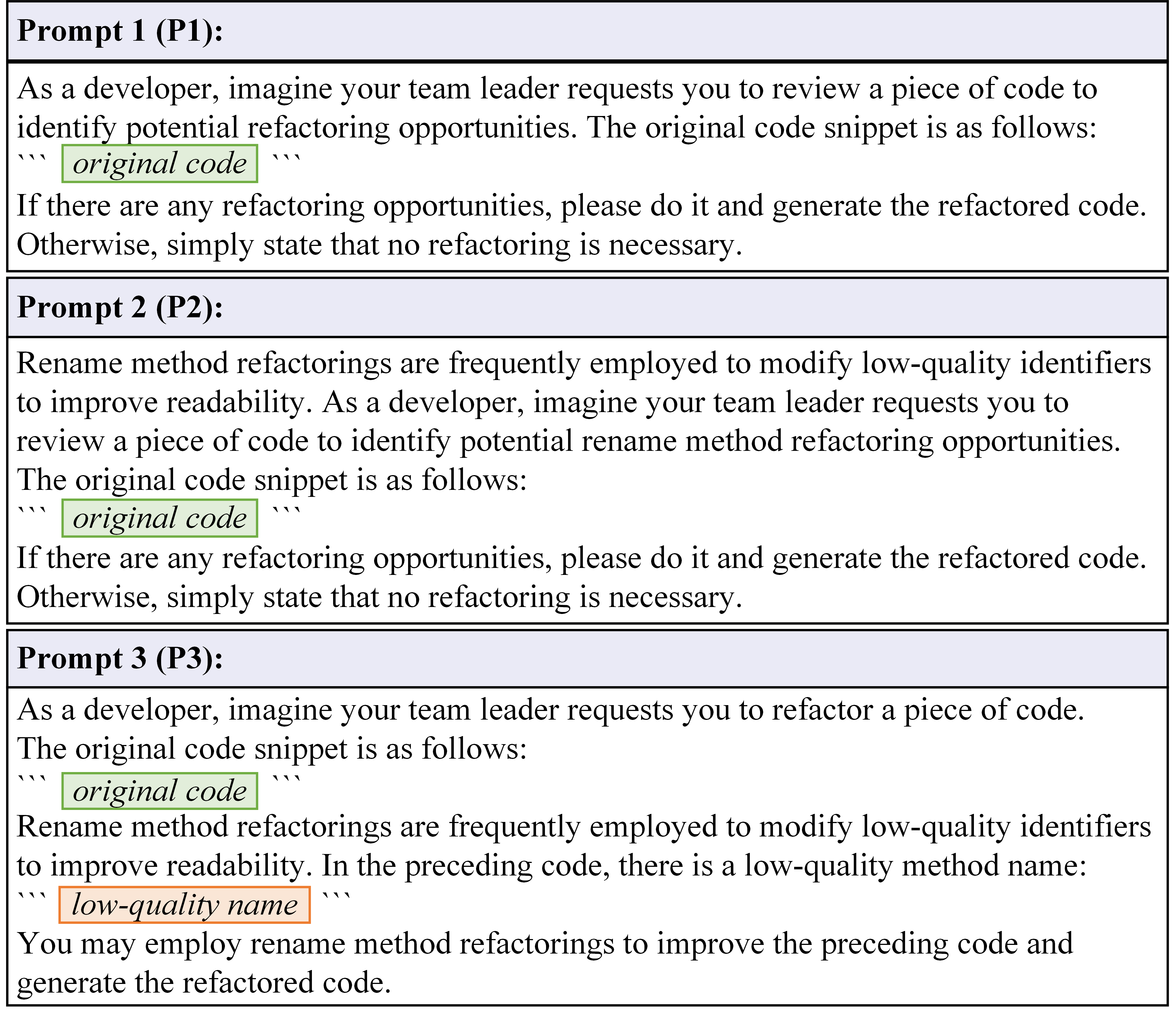}}
    \caption{Prompt Templates}
    \label{fig:prompt1}
\end{figure}

The quality of prompts could significantly influence the performance of LLMs~\cite{white2023prompt,white2024chatgpt}. To select suitable prompts, we followed well-established best practices~\cite{awesome2024prompt,prompt2024prompt} which suggests that prompts could consist of \emph{instruction}, \emph{context}, \emph{input data}, and \emph{output indicator} for better results. Moreover, in a recent study, Guo et al.~\cite{guo2023exploring} explored the potential of ChatGPT in automated code refinement and designed prompt templates that could maximize the performance of ChatGPT. Their findings suggest that prompts with concise scenario descriptions tend to generate better results. To this end, in this paper, we followed the best practices and tailored different prompt templates to answer the different research questions (as shown in Fig.~\ref{fig:prompt1}):
\begin{enumerate}
    \item To answer RQ1-1, we designed the first prompt template \textbf{P1}. It specifies the role of the large language model, the task, and the source code to be refactored.
    \item To answer RQ1-2, we designed the second prompt template \textbf{P2}. Besides all the information specified in P1, P2 also specifies the requested refactoring type. Notably, we designed a unique version of P2 for each refactoring type. 
    \item To answer RQ2, we designed the third prompt template \textbf{P3}. It specifies the role of the large language model, the task, the code entities that should be refactored, and the requested refactoring type as well as its explanation. 
\end{enumerate}

Notably, the \emph{original code} in the templates refers to the document (file) where the refactoring is expected to be done. As we explained in Section~\ref{section:dataset}, all of the selected refactorings are within-document refactoring whose key modifications are often limited to a single document. Although source code outside the enclosing document may influence the refactoring and some modifications could happen to other documents, it is impractical to input all such contexts into the prompt because of the length limitation of the LLMs. For example, a \emph{rename method} refactoring could be influenced by references (including those outside the enclosing document) to the method, and it should also update such references accordingly. However, its references could appear in many documents, and it is challenging to include all these documents in the prompt.  

\subsection{Metrics}
\label{section:label}
To evaluate the accuracy of LLMs in identifying refactoring opportunities, we define specific heuristics for each refactoring type. For a refactoring applied to a single code entity, like \emph{rename method} and \emph{inline variable} refactorings, if the conducted refactoring and the suggested refactoring are applied to the same entity (e.g., renaming the same method or inlining the same variable), we say the model has identified the refactoring opportunity. For a refactoring that involves a range of code entities, like \emph{extract method} and \emph{extract class} refactorings, if the suggested refactoring is applied to the same enclosing entity (i.e., the decomposed long method for extract method refactoring, and the decomposed large class for extract class refactoring) and they modify some common lines of source code (e.g., they extract some common statements or common code entities), we say the model succeeds in identifying the refactoring opportunity. 
Specifically, we employ well-known and widely-used metric \emph{dice coefficient}~\cite{dice1945measures,fluri2007change,falleri2014fine}, defined as:
\begin{equation}
    tolerance = \frac{2 \times \#commons}{\#extracted + \#oracle}
\end{equation}
, where $\#commons$ indicates the number of statements (code entities) extracted by LLMs are common with the refactoring oracle, to evaluate whether LLMs identify \emph{extract method} and \emph{extract class} refactoring opportunities. If the \emph{tolerance} is greater than or equal to 0.5, i.e., more than half of the extracted statements (code entities) align with the refactoring oracle's suggestions, we consider that the model successfully identifies the refactoring opportunity. Otherwise, the model is deemed to fail in identifying the refactoring opportunity.

To evaluate the quality of refactorings conducted by LLMs, we requested three experienced developers (noted as $PTC_B$) to assess the quality of the refactorings conducted by LLMs. Note that $PTC_B$ had no overlap with the participants in Section~\ref{section:dataset}. All of the participants in $PTC_B$ had rich Java experience and were familiar with software refactoring. They had a median of 10 years of programming experience, and 4.5 years working as professional software developers. Given the original code and the refactored version (i.e., the output of LLMs), each participant should depict the quality as one of the five categories:

\begin{itemize}
    \item Excellent: better than developers, score=4;
    \item Good: comparable (or identical) to developers, score=3;
    \item Poor: inferior to developers, score=2;
    \item Failed: no refactoring at all, score=1;
    \item Buggy: resulting in semantic or syntactic bugs, score=0. 
\end{itemize} 

Notably, if the solution is buggy (i.e., changing the functionality of the source code or resulting in compilation errors), we requested the participants to mark it as \emph{buggy} even if the solution could be excellent (or good) when the bug is fixed. All participants rated independently, and they reached a high consistency by achieving a Fleiss’ kappa coefficient~\cite{fleiss1971measuring} of 0.82. For each solution, if any of the three participants rated it as \emph{buggy}, we requested them to discuss it together and reach a consensus. We took the median of the three ratings (of the same refactoring solution) as its final rating.

\section{Identification of Refactoring Opportunities}
\label{section:RQ1}
\subsection{RQ1-1: Identification with Generic Commands} To answer RQ1-1, we requested GPT and Gemini to refactor the given original code using prompt template P1 introduced in Section~\ref{section:setup} as a cue. The evaluation results are presented in Table~\ref{tab1} where $\#$Succeeded and $\#$Missed represent the number of successfully identified refactoring opportunities and the number of missed refactoring opportunities, respectively. The numbers outside parentheses are absolute numbers whereas the percentages within parentheses present the ratio (i.e., Recall, also termed the Success Rate) of the number of succeeded cases to the total number of involved cases.

From Table~\ref{tab1}, we observe that GPT and Gemini successfully identified 28 and 7 out of the 180 refactoring opportunities, respectively. 
As a result, their success rates were 15.6\%=28/180 and 3.9\%=7/180, respectively. That is, they missed many more refactoring opportunities than what they successfully identified. It suggests that LLM-based identification of refactoring opportunities with generic commands may not be fruitful. 

\begin{table}
    \centering
    \caption{LLM-based Identification of Refactoring Opportunities}
    \begin{tabular}{l|rr|rr}
    \toprule
    \multirow{2}{*}{\textbf{Refactoring Type}} & \multicolumn{2}{c}{\textbf{GPT}} & \multicolumn{2}{c}{\textbf{Gemini}} \\ \cmidrule{2-5} & \textbf{\#Succeeded} & \textbf{\#Missed} & \textbf{\#Succeeded} & \textbf{\#Missed} \\
    \midrule
    Extract Class & 0 (0\%) & 20  & 0 (0\%) & 20 \\
    Extract Method & 9 (45\%) & 11 & 1 (5\%) & 19 \\
    Extract Variable & 5 (25\%) & 15 & 0 (0\%) & 20 \\
    Inline Method & 3 (15\%) & 17 & 3 (15\%) & 17 \\
    Inline Variable & 3 (15\%) & 17 & 1 (5\%) & 19 \\
    Rename Attribute & 1 (5\%) & 19 & 0 (0\%) & 20 \\
    Rename Method & 2 (10\%) & 18 & 0 (0\%) & 20 \\
    Rename Parameter & 3 (15\%) & 17 & 0 (0\%) & 20 \\
    Rename Variable & 2 (10\%) & 18  & 2 (10\%) & 18 \\
    \textbf{Total} & \textbf{28 (15.6\%)} & \textbf{152} & \textbf{7 (3.9\%)} & \textbf{173} \\
    \bottomrule
    \end{tabular}
    \label{tab1}
\end{table}

We further analyzed the evaluation results by considering the performance on different refactoring types (as shown in Table~\ref{tab1}). Our analysis results suggest that:
\begin{itemize}[listparindent=-0.5cm,leftmargin=1cm,topsep=0.06cm]
    \item Neither of them identified any of the 20 extract class refactoring opportunities, suggesting that it is rather challenging for LLMs to identify this kind of refactoring opportunities.
    \item GPT achieved a success rate of 45\% in identifying extract method refactoring opportunities. It is also the highest success rate on the table. However, we also notice that the success rate of Gemini on the same refactoring type is as low as 5\%. It may suggest that not all LLMs are good at identifying extract method refactoring opportunities. 
   \item Gemini failed to outperform GPT on any of the refactoring types. On six out of the nine refactoring types, GPT substantially outperformed Gemini. On the other three refactoring types (i.e., extract class, inline method, and rename variable), they fought to a draw. As a result, GPT identified 28 refactoring opportunities in total, 4=28/7 times the number identified by Gemini.    
\end{itemize}

To reveal what kind of changes LLMs had conducted and why they frequently failed to identify the given refactoring opportunities, we retrieved three examples for each of the nine refactoring types from the refactoring results of GPT and Gemini, resulting in a total of 54 examples (2 LLMs $\times$ 9 refactoring types $\times$ 3 examples). We manually analyzed these examples, and our evaluation results suggest that from the 54 sampled examples, LLMs identified a total of 112 refactoring opportunities, including 80 opportunities of the nine refactoring types. 76.8\%= 86/112 (i.e., accuracy) of the suggested refactoring opportunities were manually confirmed as beneficial whereas others were denied. Further analysis on the 80 refactoring opportunities suggests that:
\begin{itemize}[listparindent=-0.5cm,leftmargin=1cm,topsep=0.06cm]
    \item 72.5\%=58/80 of them were associated with extract-related refactorings (especially extract method refactorings), 22.5\%=18/80  were associated with rename-related refactorings, and 5\%=4/80  were associated with inline-related refactorings. The distribution also explains why LLMs frequently failed to identify the given types of refactoring opportunities except for extract method refactoring opportunities. 
    \item 60.3\%=35/58 of the extract-related suggestions were employed to alleviate duplicate code and the remainder focused on decomposing large classes/methods or extracting expressions (code snippets). In total, 72.4\%=42/58 of such refactoring opportunities were manually confirmed as beneficial whereas others were denied. 
    \item 72.2\%=13/18 of the rename-related refactorings suggested employing more descriptive names whereas the others suggested maintaining consistent naming styles by renaming. In total, 61.1\%=11/18 of such refactoring opportunities were manually confirmed as beneficial whereas others were denied. 
\end{itemize}

Besides the nine refactoring types, LLMs also suggested 32 refactoring opportunities of other types, e.g., \textit{simplifying code structures} and \textit{leveraging Java's new features}. For example, LLMs suggested replacing loop statements with stream API to improve code readability and running efficiency. Notably, such kinds of refactorings are less popular than the nine refactoring types investigated in this paper and some of them may not yet supported by mainstream refactoring tools, which may indicate that LLMs have the potential to suggest refactoring opportunities of less famous refactoring types. In total, 25 out of the 32 refactoring opportunities were manually confirmed as beneficial, with an accuracy of 78.1\%=25/32.

\begin{tcolorbox}[
  enhanced,
  rounded corners=all,
  colback=gray!10,
  colframe=gray!100!black,
  colbacktitle=gray!90,
  title=Answer to RQ1-1,
  fonttitle=\bfseries,
  shadow={1mm}{-1mm}{0mm}{black!50!white},
  boxrule=0.5mm,
]
LLMs exhibit limited effectiveness in identifying refactoring opportunities with general prompts, with overall low success rates. However, GPT shows a notably higher capability in identifying \emph{extract method} refactorings.
\end{tcolorbox}

\subsection{RQ1-2: Explicitly Specifying Expected Refactoring Types}
\label{sub:RQ1-2}
To investigate the performance of LLMs in identifying refactoring opportunities when expected refactoring types are explicitly specified, we requested GPT and Gemini to refactor the given original code using prompt template P2 as a cue. As specified in Section~\ref{section:label}, we manually analyzed the refactorings conducted by the LLMs, and computed the performance metrics accordingly. 

The evaluation results are presented in Table~\ref{tab2}. By comparing Table~\ref{tab1} and Table~\ref{tab2}, we observe that explicitly specifying the expected refactoring types can substantially improve the success rate in LLM-based identification of refactoring opportunities. The success rate of GPT increased substantially from 15.6\% to 52.2\%, with a relative improvement of 234.6\%=(52.2\%-15.6\%)/15.6\%. Gemini also achieved a substantial improvement of 441\%= (21.1\%-3.9\%)/3.9\%. 
We performed the Wilcoxon signed-rank test~\cite{wilcoxon1945individual} and used Cliff's Delta ($d$) as the effect size~\cite{grissom2005effect} to validate whether explicitly specifying expected refactoring types significantly improved the success rate. The test results ($p$-value=3.06E-15, Cliff's $|d|$=0.37) confirmed that the improvement of GPT was statistically significant. Similarly, the improvement of Gemini was statistically significant ($p$-value=1.61E-7 and Cliff's $|d|$=0.17).

\begin{table}
    \centering
    \caption{LLM-based Identification of Refactoring Opportunities (Specifying Expected Refactoring Types)}
    \begin{tabular}{l|rr|rrr}
    \toprule
    \multirow{2}{*}{\textbf{Refactoring Type}} & \multicolumn{2}{c}{\textbf{GPT}} & \multicolumn{2}{c}{\textbf{Gemini}} \\ \cmidrule{2-5} & \textbf{\#Succeeded} & \textbf{\#Missed} & \textbf{\#Succeeded} & \textbf{\#Missed} \\
    \midrule
    Extract Class & 9 (45\%) & 11 & 2 (10\%) & 18 \\
    Extract Method & 12 (60\%) & 8 & 5 (25\%) & 15 \\
    Extract Variable & 7 (35\%) & 13 & 1 (5\%) & 19 \\
    Inline Method & 7 (35\%) & 13 & 6 (30\%) & 14\\
    Inline Variable & 5 (25\%) & 15 & 3 (15\%) & 17 \\
    Rename Attribute & 14 (70\%) & 6 & 8 (40\%) & 12 \\
    Rename Method & 12 (60\%) & 8 & 6 (30\%) & 14 \\
    Rename Parameter & 14 (70\%) & 6 & 4 (20\%) & 16 \\
    Rename Variable & 14 (70\%) & 6 & 3 (15\%) & 17 \\
    \textbf{Total} & \textbf{94 (52.2\%)} & \textbf{86} & \textbf{38 (21.1\%)} & \textbf{142} \\
    \bottomrule
    \end{tabular}
    \label{tab2}
\end{table}

From Table~\ref{tab2}, we also observe that the success rate varies substantially among different refactoring types:
\begin{itemize}[listparindent=-0.5cm,leftmargin=1cm,topsep=0.06cm]
    \item Explicitly specifying the expected refactoring types did not reduce LLMs' success rate on any refactoring types.  
    \item Explicitly specifying the expected refactoring types improved the success rate of GPT on all types of refactoring opportunities whereas it only improved Gemini on four out of the nine categories of refactoring opportunities, i.e., extract method, rename attribute, rename method, and rename parameter.
    \item The success rate of GPT varies from 25\% to 70\% whereas the success rate of Gemini varies from 5\% to 40\%. It may suggest that LLM-based refactoring recommendation could be more fruitful on some refactoring types.
    \item GPT achieved a high success rate of 67.5\%=(54/80) in identifying rename refactoring opportunities, including rename attribute, rename method, rename parameter, and rename variable. However, we also observe that Gemini's success rate on rename refactoring is not substantially higher than that on other refactorings. It may suggest that different large language models could be good at identifying different types of refactoring opportunities, and thus they are potentially complementary.
    \item The success rate on inline refactoring opportunities (i.e., inline method and inline variable) remains low even if the expected refactoring types are explicitly specified.  
\end{itemize}


When the expected refactoring types are explicitly specified, we observe that LLMs significantly increase the number of identified refactoring opportunities. GPT achieved an improvement of 235.7\%=(94-28)/28, while Gemini achieved an improvement of 442.9\%=(38-7)/7. We leverage the example in Listing~\ref{listing:success} to illustrate that LLMs can accurately identify refactoring opportunities that are consistent with those actually conducted by developers.
In this example, the original developer extracted the conditional expression in the \emph{if} statement (Line 7) into the variable (Line 5) to improve the readability and maintainability of the code. GPT also accurately identified this refactoring opportunity and conducted the same \emph{extract variable} refactoring as conducted by the developer. Therefore, we unanimously believe that GPT successfully identified a refactoring opportunity (i.e., \#Succeeded).

\begin{lstlisting}[float, language=diff, numbers=left, xleftmargin=10em, xrightmargin=10em, frame=shadowbox, caption={Refactoring Opportunity Successfully Identified by LLM}, breaklines=true, label={listing:success}, belowskip=-1.0 \baselineskip]
   /* From DefaultRefreshEventListener.java in Hibernate-ORM */
   public void onRefresh(RefreshEvent event, RefreshContext refreshedAlready) {
       ......
       if (persistenceContext.reassociateIfUninitializedProxy(object)) {
+         %*\colorbox{codegreen}{boolean isTransient = isTransient(event, source, object);}*) 
           ......
-         if (%*\colorbox{codered}{isTransient(event, source, object)}*)) {
+         if (%*\colorbox{codegreen}{isTransient}*)) {
              source.setReadOnly(object, source.isDefaultReadOnly());
           }
       ......
   }
\end{lstlisting}

To compute the accuracy in identifying refactoring opportunities, we recruited three evaluators (i.e., $PTC_B$ as introduced in Section~\ref{section:label}) to manually validate all refactorings conducted by LLMs. If LLMs conducted a refactoring of the expected refactoring type and this refactoring was not found in the ground truth, it is a false positive if at least two of the three evaluators deemed it \textit{unnecessary}. Note that a refactoring is unnecessary if it cannot improve the quality of the source code. The evaluation results suggest that both GPT and Gemini reported some false positives, which had a negative impact on their accuracy. For example, GPT suggested 731 refactorings of the expected types, of which 501 were considered necessary and 230 were denied by human experts. As a result, the accuracy of GPT is 68.5\%=501/731. Gemini achieved a comparable accuracy of 64.9\%=61/94.

\begin{tcolorbox}[
  enhanced,
  rounded corners=all,
  colback=gray!10,
  colframe=gray!100!black,
  colbacktitle=gray!90,
  title=Answer to RQ1-2,
  fonttitle=\bfseries,
  shadow={1mm}{-1mm}{0mm}{black!50!white},
  boxrule=0.5mm,
]
When refactoring types are explicitly specified, the success rate of LLMs increases significantly. However, some refactoring types, especially \emph{inline} refactorings, still have a low success rate.
\end{tcolorbox}

\subsection{RQ1-3: Size of Source Code}
\label{sub:RQ1-3}
To investigate how the size of the source code influences the success rate, we calculated the point-biserial correlation coefficients~\cite{tate1954correlation} between the size of the to-be-refactored source code and the success rate of LLMs in identifying refactoring opportunities. We measured the size of the source code with the well-known metric LOC (lines of code). 

\begin{figure}[t]
    \centerline{\includegraphics[width=0.62\linewidth]{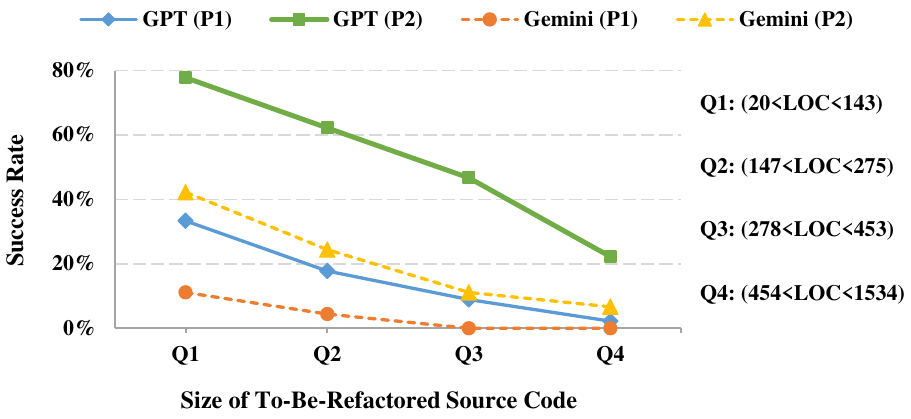}}
    \caption{Size of Source Code Influences LLMs' Success Rate}
    \label{fig:line}
\end{figure}

Our computation results suggest that when prompt template P2 was used, there is a moderate (GPT, -0.38) or weak (Gemini, -0.28) negative correlation with the size of the to-be-refactored source code: The larger the source code is, the lower the success rate is. When we switched from prompt template P2 to P1 (without explicitly specifying the expected refactoring types), the negative correlation remained (-0.29 for GPT and -0.2 for Gemini).
Besides the correlation coefficient, we also categorized the source code into four equally-sized groups according to their size (LOC), noted as  Q1, Q2, Q3, and Q4 where Q1 was composed of the 25\% of the smallest code snippets. The evaluation results are presented in Fig.~\ref{fig:line}.

The horizontal axis outlines the four groups, along with the minimum and maximum sizes of LOC within the given group. The vertical axis represents the success rate in refactoring opportunity identification. We observe from this figure that the success rate decreased quickly with the increase in LOC.

The finding, i.e., it is more challenging to identify the given refactoring opportunities accurately from longer to-be-refactoring source code, is reasonable. It is well-known that the longer the prompt is, the more challenging it is for LLMs to understand the input and generate expected answers~\cite{chang2023prompt}. Consequently, when the to-be-refactored source code is lengthy and complex, it is reasonable that LLM-based identification of refactoring opportunities could be less successful.
We also observe that GPT identified more refactoring opportunities than Gemini under the same prompt. Furthermore, Gemini demonstrates a higher success rate in identifying refactoring opportunities when the proper prompts are given (i.e., P2), compared to GPT when the refactoring types are not explicitly specified (i.e., P1). The finding highlights the importance of prompt settings in maximizing the potential of LLMs in identifying refactoring opportunities.

\begin{tcolorbox}[
  enhanced,
  rounded corners=all,
  colback=gray!10,
  colframe=gray!100!black,
  colbacktitle=gray!90,
  title=Answer to RQ1-3,
  fonttitle=\bfseries,
  shadow={1mm}{-1mm}{0mm}{black!50!white},
  boxrule=0.5mm,
]
The size of the source code negatively influences the success rate of LLMs in identifying refactoring opportunities, with longer source code resulting in lower success rates.
\end{tcolorbox}

\subsection{RQ1-4: Strengths and Weakness of LLMs}
\label{sub:RQ1-4}
To gain a deeper understanding of the strengths and weaknesses of LLM-based identification of refactoring opportunities, we manually analyzed the reasons behind the refactorings in the subject dataset and classified them into 23 refactoring subcategories (reasons). For example, \emph{code duplication}, one of the most common code smells, refers to the duplicated code snippets within a software project~\cite{fowler1999refactoring,yamashita2013developers}. The \emph{extract method} refactoring is an effective and commonly employed strategy to mitigate this issue~\cite{silva2016we,alomar2024behind}.
\begin{figure}[t]
    \centerline{\includegraphics[width=1.0\linewidth]{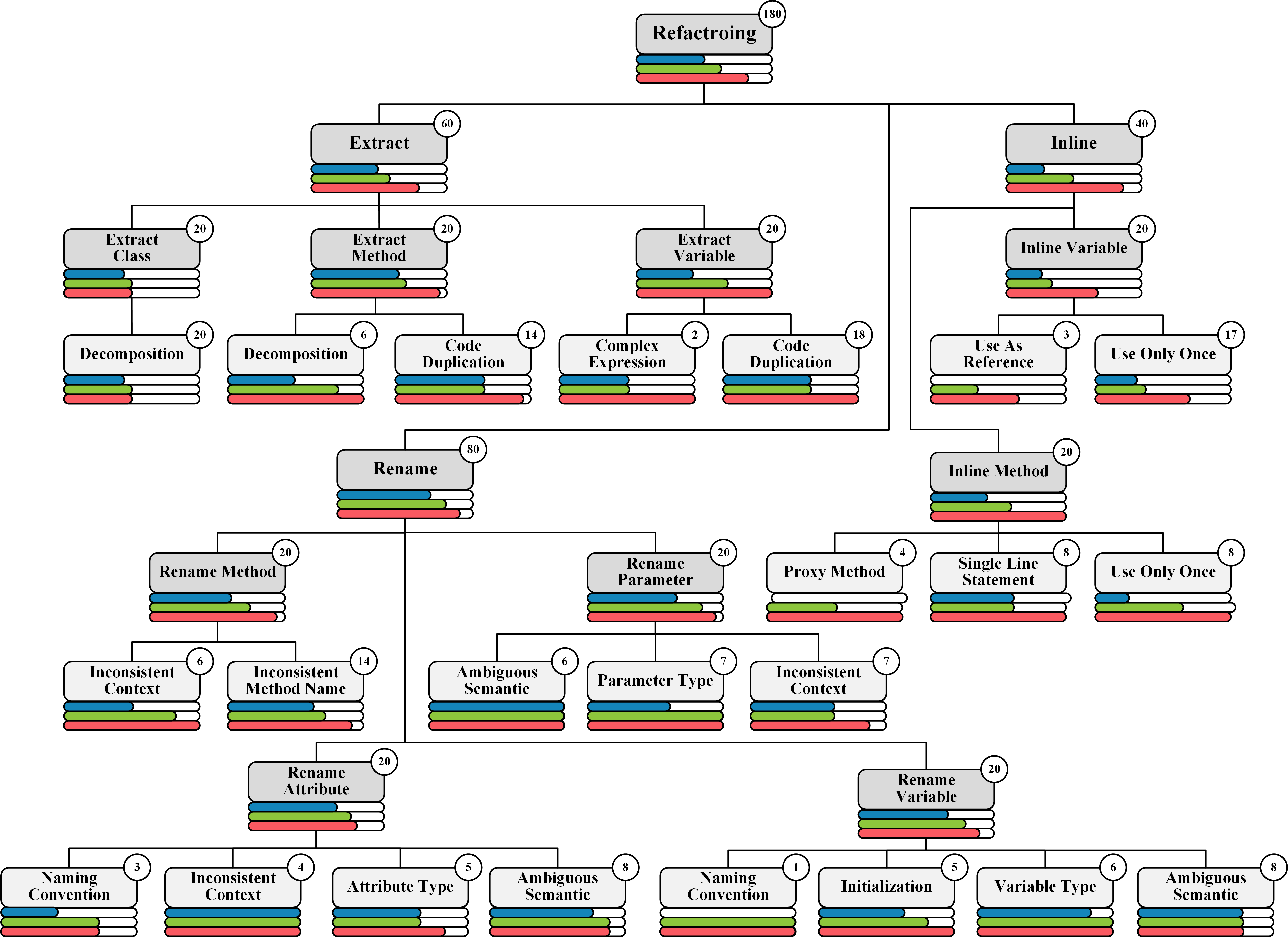}}
    \caption{Taxonomy of refactoring subcategories for automated software refactoring. We present the percentage of successful identification for each refactoring subcategory with bars below the corresponding category: P2 (blue bar), P2 + Refactoring Subcategory (green bar), P2 + Refactoring Subcategory + Search Space Limitation (red bar).}
    \label{fig:taxonomy}
\end{figure}
Fig.~\ref{fig:taxonomy} illustrates the taxonomy of these subcategories. The number on the top-right corner of each label in this figure represents the number of instances manually assigned to each refactoring subcategory (e.g., 14 refers to \emph{inconsistent method name}).
In the previous research questions (RQs), we observe that GPT consistently outperformed Gemini across all types of refactorings and in both prompt settings. Consequently, this section concentrates on cases where the state-of-the-art LLM (i.e., GPT) performed well or not, and explores potential improvements in its capability to identify refactoring opportunities.

\begin{figure}[t]
    \centerline{\includegraphics[width=0.7\linewidth]{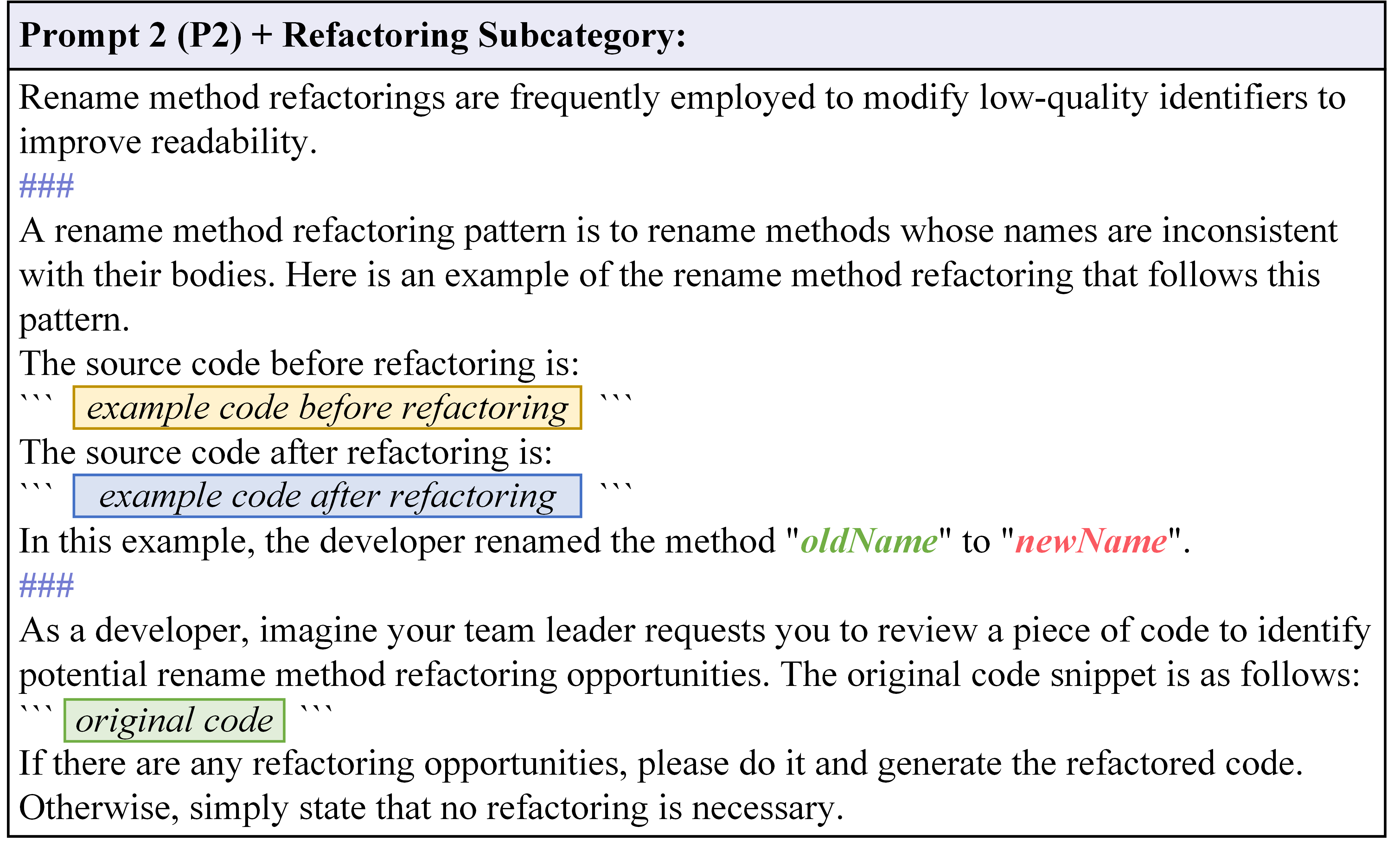}}
    \caption{P2 + Refactoring Subcategory}
    \label{fig:promptandsubcategory}
\end{figure}

The blue bar represents the success rate of GPT in identifying refactoring opportunities under P2. An empty bar represents that GPT always failed in that category. From Fig.~\ref{fig:taxonomy}, 
we make the following observations:
\begin{itemize}[listparindent=-0.5cm,leftmargin=1cm,topsep=0.06cm]
    \item GPT performed poorly in identifying inline-related refactoring opportunities, especially for variables that are only references to other objects (\emph{use as reference}) and methods that are only called by a proxy method (\emph{proxy method}). In both cases, GPT failed to identify any refactoring opportunities, suggesting that GPT does not excel at identifying such refactoring opportunities.
    \item For extract-related refactorings, the success rates did not significantly vary across subcategories. One possible explanation is that GPT possesses a robust understanding of the motives categorized for these refactorings.
    We also observe that GPT showed greater proficiency in removing code duplication compared to decomposing classes and methods as well as extracting complex expressions.
    \item Although the reasons for rename-related refactoring are more diverse than the other two types of refactorings, GPT showed the highest success rate in identifying such refactoring opportunities. 
    However, the effectiveness of GPT varied for refactoring subcategories. For example, it rarely identified refactoring opportunities associated with \emph{naming conventions}, 
    whereas it had a higher success rate in identifying refactoring opportunities for \emph{ambiguous semantics}. 
\end{itemize}
To further investigate the performance of GPT in identifying refactoring opportunities across different subcategories (e.g., \emph{code duplication} and \emph{naming convention}), 
we designed a new prompt template \textbf{P2 + Refactoring Subcategory} for each refactoring subcategory, as illustrated in Fig.~\ref{fig:promptandsubcategory}.

With the increasing ability of LLMs, in-context learning has become more widely adopted. By providing a few examples of desired inputs and outputs within the context, 
LLMs can better understand and adapt to specific tasks (e.g., refactoring opportunity identification), enhancing their effectiveness. To this end, we retrieve similar refactoring patterns from the history of the subject projects to serve as contextual cues, instructing the LLMs to understand the necessary background knowledge for each refactoring subcategory.
Notably, the retrieved examples can be used repeatedly because the refactoring patterns corresponding to each refactoring subcategory are consistent. For example, expressions that are repeatly used should be extrated as a new variable to enhance code readability and maintainability.

The full prompt templates are available in our replication package~\cite{replication2024package}.
As presented in Fig.~\ref{fig:taxonomy}, the green bar represents the success rate of GPT in identifying refactoring opportunities under P2 + Refactoring Subcategory.
We observe from this figure that explicitly specifying the expected refactoring Subcategories can improve the success rate in LLM-based identification of refactoring opportunities. 
Specifically, the success rate of GPT increased significantly from 52.2\% to 66.7\%, with a relative improvement of 27.8\%=(66.7\%-52.2\%)/52.2\%. Furthermore, there was no observed decrease in success rate across any subcategory when refactoring subcategories were explicitly specified.
We performed the same significance test as described in Section~\ref{sub:RQ1-2} to validate whether explicitly specifying expected refactoring Subcategories significantly improved the success rate. The test results ($p$-value=1.26E-4, Cliff's $|d|$=0.14) confirmed that the improvement was statistically significant.

We conclude in Section~\ref{sub:RQ1-3} that the success rate of identifying refactoring opportunities is affected by the size of the source code. The finding highlights the challenge of LLMs in effectively understanding and handling longer and more complex source code.
Consequently, it may not be a wise choice to feed the entire document to LLMs for specific refactoring types. For example, \emph{extract variable} refactorings typically do not involve code snippets beyond the scope of the enclosing method of the extracted expressions. To this end, we tried to narrow the search space based on P2 + Refactoring Subcategory. Notably, for each refactoring subcategory, we designed a unique strategy to narrow the search space, noted as \textbf{Search Space Limitation}. 
In the example illustrated in Fig.~\ref{fig:promptandsubcategory}, the search space was limited from the entire document to a single method because the method name is only associated with the functionality of its method body. The full strategies are detailed in our replication package~\cite{replication2024package}.
As presented in Fig.~\ref{fig:taxonomy}, the red bar represents the success rate of GPT in identifying refactoring opportunities under P2 + Refactoring Subcategory + Search Space Limitation.
From this figure, we observe that narrowing the search space significantly boosts the success rate of GPT in identifying refactoring opportunities from 66.7\% to 86.7\%, with a relative improvement of 30\%=(86.7\%-66.7\%)/66.7\%.
Moreover, the proposed strategy did not reduce the performance of GPT in any refactoring subcategory.

\begin{table*}
    \centering
    \caption{LLM-based Identification of Refactoring Opportunities (Specifying Expected Refactoring Subcategories ($\$$) And Narrowing the Search Space ($\dagger$))}
    \begin{tabular}{l|rr|rr|rr}
    \toprule
    \multirow{2}{*}{\textbf{Refactoring Type}} & \multicolumn{2}{c}{\textbf{P2}} & \multicolumn{2}{c}{\textbf{P2 $^\$$}} & \multicolumn{2}{c}{\textbf{P2 $^{\$\,\dagger}$}} 
    \\ \cmidrule{2-7} & \textbf{\#Succeeded} & \textbf{\#Missed} & \textbf{\#Succeeded} & \textbf{\#Missed} & \textbf{\#Succeeded} & \textbf{\#Missed} \\
    \midrule
    Extract Class & 9 (45\%) & 11 & 10 (50\%) & 10 & 10 (50\%) & 10 \\
    Extract Method & 12 (60\%) & 8 & 14 (70\%) & 6 & 19 (95\%) & 1 \\
    Extract Variable & 7 (35\%) & 13 & 13 (65\%) & 7 & 20 (100\%) & 0 \\
    Inline Method & 7 (35\%) & 13 & 12 (60\%) & 8 & 20 (100\%) & 0 \\
    Inline Variable & 5 (25\%) & 15 & 7 (35\%) & 13 & 14 (70\%) & 6 \\
    Rename Attribute & 14 (70\%) & 6 & 16 (80\%) & 4 & 17 (85\%) & 3 \\
    Rename Method & 12 (60\%) & 8 & 15 (75\%) & 5 & 19 (95\%) & 1 \\
    Rename Parameter & 14 (70\%) & 6 & 17 (85\%) & 3 & 19 (95\%) & 1 \\
    Rename Variable & 14 (70\%) & 6 & 16 (80\%) & 4 & 18 (90\%) & 2  \\
    \textbf{Total} & \textbf{94 (52.2\%)} & \textbf{86} & \textbf{120 (66.7\%)} & \textbf{60} & \textbf{156 (86.7\%)} & \textbf{24} \\
    \bottomrule
    \end{tabular}
    \label{tab3}
\end{table*}

Table~\ref{tab3} presents the performance of GPT in identifying nine types of refactoring opportunities refactoring subcategories are explicitly specified and the search space is narrowed.
From this table, we make the following observations:
\begin{itemize}[listparindent=-0.5cm,leftmargin=1cm,topsep=0.06cm]
    \item For \emph{extract variable} and \emph{inline method} refactorings, GPT successfully identified all of the refactoring opportunities. Notably, under the original P2, 
    the success rate of GPT in identifying these two types of refactoring opportunities was only 35\%, which may reveal that the proposed two strategies: ``Refactoring Subcategory'' and ``Search Space Limitation'' are both effective.
    \item Similarly, the application of these strategies significantly improved the identification of refactoring opportunities for \emph{extract method}, \emph{rename method}, \emph{rename parameter}, and \emph{rename variable} refactorings.
    \item The proposed strategies sometimes did not yield significant improvements for \emph{extract class} and \emph{rename attribute} refactorings, which reveals that 
    GPT's capability in identifying such refactorings may be limited.
    \item Although the proposed strategies significantly improved GPT's success rate in identifying \emph{inline variable} refactoring opportunities, its performance still requires further improvement. 
    For \emph{extract class} refactoring, the performance of GPT remains unsatisfactory. Therefore, we suggest that more efforts should be invested in the future on how to exploit LLMs to more accurately identify such refactoring opportunities.
\end{itemize}

We also performed the significance test to validate whether narrowing the search space significantly improved the success rate. The test results ($p$-value=5.22E-9, Cliff's $|d|$=0.2) confirmed that the improvement was statistically significant.
Although the proposed strategies improve the capability of LLMs to identify refactoring opportunities, some cases still pose challenges for accurate identification. We illustrate the possible reasons with the example in Listing~\ref{listing:miss}, where the original developer renamed the attribute ``\emph{MAX$\_$CAPACITY}" to ``\emph{MIN$\_$CAPACITY}" to better align with the underlying code intent, as described in~\cite{jira2024hadoop}. However, in this example, GPT always failed to identify this refactoring opportunity. Under the optimal prompt setting, GPT described ``\emph{MAX$\_$CAPACITY}" as ``\emph{A clear constant name indicating the maximum capacity of the cache.}", which suggests that it believed the original name aptly expressed the attribute's functionality.
The failure of the LLMs to identify this refactoring opportunity may stem from several reasons:
\begin{lstlisting}[float, language=diff, numbers=left, xleftmargin=9em,xrightmargin=9em, caption={Refactoring Opportunity Missed by LLM}, breaklines=true, label={listing:miss}, belowskip=-1.0 \baselineskip]
   /* From RetryCache.java in Hadoop */
   public class RetryCache {
       public static final Logger LOG = LoggerFactory.getLogger(RetryCache.class);
       private final RetryCacheMetrics retryCacheMetrics;
-     private static final int %*\colorbox{codered}{MAX\_CAPACITY}*) = 16;
+     private static final int %*\colorbox{codegreen}{MIN\_CAPACITY}*) = 16;
       ......
       public RetryCache(String cacheName, double percentage, long expirationTime)   {
           int capacity = LightWeightGSet.computeCapacity(percentage, cacheName);
-         capacity = Math.max(capacity, %*\colorbox{codered}{MAX\_CAPACITY}*));
+         capacity = Math.max(capacity, %*\colorbox{codegreen}{MIN\_CAPACITY}*));           
           ......
        }
   }
\end{lstlisting}
\begin{itemize}[listparindent=-0.5cm,leftmargin=1cm,topsep=0.06cm]
    \item \textbf{Literal Interpretation:} The name ``\emph{MAX$\_$CAPACITY}" straightforwardly suggests the maximum capacity. Without explicit context suggesting that it should represent the minimum capacity, LLMs may not deduce the need for renaming.
    \item \textbf{Complexity of Code Intent:} ``\emph{MAX$\_$CAPACITY}" serves as a lower bound in capacity calculations and may exceed the LLMs' inferential capabilities, which involves understanding the role of functions \texttt{Math.max} in that specific context.
    \item \textbf{Limitation of Training Data:} If the LLMs' training data lacks similar contexts or if such refactoring patterns are rare, the models may not generalize well to the specific scenario, leading to difficulty in identifying such refactoring opportunities.
    \item \textbf{Lack of Domain-Specific Knowledge:} Despite their broad programming knowledge, LLMs may not possess detailed insights into certain best practices or naming conventions essential for identifying such refactoring opportunities.
\end{itemize}

\begin{tcolorbox}[
  enhanced,
  rounded corners=all,
  colback=gray!10,
  colframe=gray!100!black,
  colbacktitle=gray!90,
  title=Answer to RQ1-4,
  fonttitle=\bfseries,
  shadow={1mm}{-1mm}{0mm}{black!50!white},
  boxrule=0.5mm,
]
LLMs perform better in areas like code duplication removal and rename-related refactorings, but struggle with class decomposition. Furthermore, specifying refactoring subcategories and narrowing the search space significantly improve their performance.
\end{tcolorbox}

\section{Recommendation of Refactoring Solutions}
\label{section:RQ2}
\subsection{RQ2-1: Quality of Refactoring Solutions}
\label{section:RQ2-1}
To investigate the quality of refactorings suggested (and conducted) by LLMs, we requested them to refactor the given original code using P3 as a cue. As a result, GPT and Gemini recommended 176 and 137 refactoring solutions for the 180 requests. 
We established a test suite to evaluate the safety of refactoring solutions suggested by LLMs. The suite comprises a total of 102 unit tests, derived from two sources: 59 are from the inherent unit tests in the subject dataset, while the remaining 43 are generated for the original code (i.e., the entire document) using an automated unit test generation tool~\cite{yaron2024testme}. Notably, the tool was selected because 1) it supports various Java versions and 2) it can generate unit tests directly on the source code.

We initially ran these unit tests on the original code before refactoring, and the results confirmed that all unit tests passed successfully. Subsequently, we executed them again on the refactoring solutions suggested by GPT and Gemini, respectively, resulting in a total of 4 refactoring solutions that failed to pass the unit tests. 
To ensure no bugs were overlooked, we also requested developers (i.e., $PTC_B$) to manually inspect both the refactoring solutions that passed the unit tests and those for which unit tests could not be generated. This dual approach, combining automated testing with manual inspection, is designed to provide a thorough evaluation of the safety of LLM-based refactoring. In addition, the developers were requested to rate the quality of the refactoring solutions.

\begin{figure}[t]
    \centerline{\includegraphics[width=0.62\linewidth]{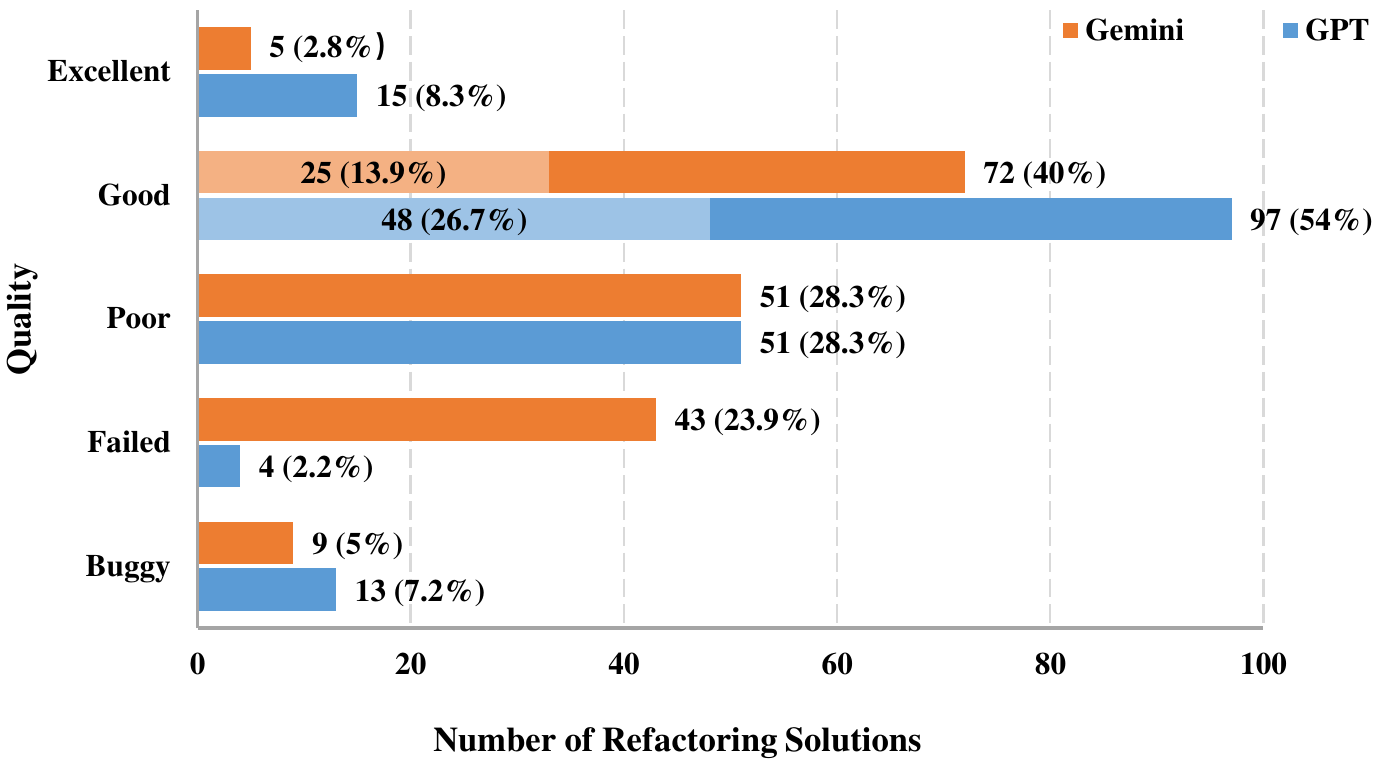}}
    \caption{Quality of LLMs' Refactoring Solutions}
    \label{fig:rq2}
\end{figure}

Fig.~\ref{fig:rq2} presents the results of human evaluation of the suggested solutions. The numbers before parentheses are the absolute numbers of solutions receiving the given rating whereas the percentages within parentheses present the ratio of the absolute numbers to the total number of requests (i.e., 180).  
From this figure, we make the following observations:
\begin{itemize}[listparindent=-0.5cm,leftmargin=1cm,topsep=0.06cm]
    \item First, LLMs have the potential in automated suggestions of high-quality refactoring solutions. For GPT, 63.6\%=(15+97)/176 of the suggested solutions are comparable to (even better than) those constructed by human experts. Although Gemini is less accurate, more than half (56.2\%=(5+72)/ 137) of its solutions are comparable to or better than manually constructed solutions. We also notice that refactoring solutions suggested by LLMs are sometimes identical to the ground truth (constructed by the original developers). 
    Out of the 176 solutions suggested by GPT and the 137 solutions suggested by Gemini, 48 and 25 solutions respectively are identical to their corresponding ground truth.
    \item Second, refactoring solutions suggested by LLMs cannot be accepted with further analysis. We notice that around one-third of the suggested refactoring solutions (29\%=51/176 for GPT and 37.2\%=51/137 for Gemini) are poor, i.e., inferior to the corresponding manually constructed refactoring solutions.    
    \item Third, LLMs may fail to suggest any solutions for the given refactorings. GPT and Gemini failed to make suggestions on 4 and 43 out of the 180 requests. 
    \item Finally, the refactoring solutions suggested by LLMs could be unsafe, i.e., buggy. 7.4\%=13/176 and 6.6\%= 9/137 of the refactoring solutions suggested by GPT and Gemini respectively are buggy. Among such 22=13+9 buggy solutions, 18 results in semantic changes and 4 results in syntax errors. 
\end{itemize}

On some cases, the overall quality of the suggested refactoring solutions is acceptable although they may result in semantic and/or syntactic bugs. That is, if the bugs are fixed, the solutions are comparable or even better than manually constructed ones. To validate this assumption, we manually fixed bugs and requested the participants to assess the refactorings again. Our evaluation results suggest that such manually fixed refactorings were frequently marked as good (14 out of 22 cases), with 2 marked as excellent and 6 marked as poor. 

\begin{table*}
    \caption{Quality of LLMs' Refactoring Solutions}
    \begin{tabular}{l|l|rrrrr}
    \toprule
    \multirow{2}{*}{\textbf{Refactoring Type}} & \multirow{2}{*}{\textbf{Model}} & \multicolumn{5}{c}{\textbf{Quality}} \\ \cmidrule{3-7}
                                      &                        & \textbf{\#Excellent} & \textbf{\#Good} & \textbf{\#Poor} & \textbf{\#Failed} & \textbf{\#Buggy} \\
    \midrule
    \multirow{2}{*}{Extract Class}    & GPT                    &   0 (0\%) & 13 (65\%) & 3 (15\%) & 3 (15\%) & 1 (5\%) \\
    & Gemini                 &  0 (0\%) & 10 (50\%) & 5 (25\%) & 4 (20\%) & 1 (5\%) \\ \midrule
    \multirow{2}{*}{Extract Method}    & GPT                    &  3 (15\%) & 10 (50\%) & 4 (20\%) & 0 (0\%) & 3 (15\%) \\
    & Gemini                 &  1 (5\%) & 9 (45\%) & 2 (10\%) & 8 (40\%) & 0 (0\%) \\ \midrule
    \multirow{2}{*}{Extract Variable}    & GPT                    &   4 (20\%) & 12 (60\%) & 1 (5\%) & 0 (0\%) & 3 (15\%) \\
    & Gemini                 &  0 (0\%) & 8 (40\%) & 4 (20\%) & 6 (30\%) & 2 (10\%) \\ \midrule
    \multirow{2}{*}{Inline Method}    & GPT                    &   1 (5\%) & 18 (90\%) & 1 (5\%) & 0 (0\%) & 0 (0\%) \\
    & Gemini                 &  0 (0\%) & 12 (60\%) & 1 (5\%) & 5 (25\%) & 2 (10\%) \\ \midrule
    \multirow{2}{*}{Inline Variable}    & GPT                    &  0 (0\%) & 16 (80\%) & 1 (5\%) & 0 (0\%) & 3 (15\%) \\
    & Gemini                 &  0 (0\%) & 13 (65\%) & 0 (0\%) & 5 (25\%) & 2 (10\%) \\ \midrule
    \multirow{2}{*}{Rename Attribute}    & GPT                    &  1 (5\%) & 6 (30\%) & 11 (55\%) & 1 (5\%) & 1 (5\%) \\
    & Gemini                 &  0 (0\%) & 7 (35\%) & 9 (45\%) & 3 (15\%) & 1 (5\%) \\ \midrule
    \multirow{2}{*}{Rename Method}    & GPT                    &  2 (10\%) & 5 (25\%) & 12 (60\%) & 0 (0\%) & 1 (5\%) \\
    & Gemini                 &  1 (5\%) & 4 (20\%) & 13 (65\%) & 2 (10\%) & 0 (0\%) \\ \midrule
    \multirow{2}{*}{Rename Parameter}    & GPT                 & 2 (10\%) & 6 (30\%) & 11 (55\%) & 0 (0\%) & 1 (5\%) \\
    & Gemini                 &  1 (5\%) & 4 (20\%) & 8 (40\%) & 6 (30\%) & 1 (5\%) \\ \midrule
    \multirow{2}{*}{Rename Variable}    & GPT                    & 2 (10\%) & 11 (55\%) & 7 (35\%) & 0 (0\%) & 0 (0\%) \\
    & Gemini                 &  2 (10\%) & 5 (25\%) & 9 (45\%) & 4 (20\%) & 0 (0\%) \\ \midrule
    \multirow{2}{*}{\textbf{Total}}    & \textbf{GPT}                    & \textbf{15 (8.3\%)} & \textbf{97 (54\%)} & \textbf{51 (28.3\%)} & \textbf{4 (2.2\%)} & \textbf{13 (7.2\%)} \\
    & \textbf{Gemini}               & \textbf{5 (2.8\%)} & \textbf{72 (40\%)} & \textbf{51 (28.3\%)} & \textbf{43 (23.9\%)} & \textbf{9 (5\%)} \\
    \bottomrule
    \end{tabular}
    \label{tab4}
\end{table*}

We further investigate how well the LLMs in suggesting solutions for different refactoring types, and the evaluation results are presented in Table~\ref{tab4}.
From this table, we make the following observations:
\begin{itemize}[listparindent=-0.5cm,leftmargin=1cm,topsep=0.06cm]
    \item  The quality of the suggested refactoring solutions varies substantially among different refactoring types. For example, the percentage of excellent GPT solutions varies from zero to 20\%. The percentage of poor Gemini solutions also varies substantially from zero to 65\%. It may suggest that it is often not equally challenging to recommend solutions for different types of refactorings. 
    \item LLMs works well in recommending solutions for inline-related refactorings (i.e., \emph{inline method} and \emph{inline variable}) and extraction-related refactorings (i.e., \emph{extract class}, \emph{extract method}, and \emph{extract variable}). We observe that 87.5\% of the solutions to inline-related refactorings are comparable to or even better than those conducted by developers, and 70\% of the solutions to extract-related refactorings are comparable to or even better than manually constructed ones. 
    \item Suggesting new names for rename refactorings is often more challenging. We notice that only 43.8\% of the refactoring solutions (i.e., new names for the to-be-renamed code entities) are comparable to or even better than manually constructed solutions. 
    \item Finally, LLMs result in buggy solutions on all refactoring types except for rename variable. 
\end{itemize}

\begin{lstlisting}[float, language=diff, numbers=left, xleftmargin=10em, xrightmargin=10em, frame=shadowbox, caption={LLM's suggested solution outperforms developer's suggestion}, breaklines=true, label={listing:excellent}, belowskip=-1.0 \baselineskip]
    /* From XmlPrinterTest.java in JavaParser */
    private void assertXMLEquals(
 -                      %*\colorbox{codered}{String xml1, String actual}*)) {  // original parameter
 +                      %*\colorbox{codegreen}{String expected, String actual}*)) {  // ground truth
 +                      %*\colorbox{codeblue}{String expectedXML, String actualXML}*)) {  // GPT' result
        final Document expectedDocument = getDocument(%*\colorbox{codered}{xml1}*));
 +                                                                   getDocument(%*\colorbox{codegreen}{expected}*));
 +                                                                   getDocument(%*\colorbox{codeblue}{expectedXML}*));
        final Document actualDocument = getDocument(%*\colorbox{codered}{actual}*));
 +                                                              getDocument(%*\colorbox{codegreen}{actual}*));
 +                                                              getDocument(%*\colorbox{codeblue}{actualXML}*));
        ...... 
    }
\end{lstlisting}

We observe that many of the refactoring solutions conducted by LLMs tend to outperform developers' solutions. An example is presented in Listing~\ref{listing:excellent}. In this example, the developer renamed the parameter ``\emph{xml}1" into ``\emph{expected}" (Lines 3-4) because the original parameter name lacked clarity and understandability within this method. GPT suggested the new name ``\emph{expectedXML}" is even better because it not only specifies the difference between the two parameters (expected vs. actual) but also specifies what it is (an XML document). The structure of the suggested name (\textit{adjective}+\textit{noun}) is also highly preferred by parameter names. We also notice that GPT suggested naming the other parameter ``\emph{actual}" into ``\emph{actualXML}".

\begin{tcolorbox}[
  enhanced,
  rounded corners=all,
  colback=gray!10,
  colframe=gray!100!black,
  colbacktitle=gray!90,
  title=Answer to RQ2-1,
  fonttitle=\bfseries,
  shadow={1mm}{-1mm}{0mm}{black!50!white},
  boxrule=0.5mm
]
LLMs demonstrate a promising potential in suggesting high-quality refactoring solutions, with over 60\% of solutions often comparable to or even better than those crafted by human experts.
\end{tcolorbox}

\subsection{RQ2-2: Safety of Suggested Refactoring Solutions}
As mentioned in the preceding section, refactoring solutions suggested by the evaluated large language models could be unsafe (buggy). Our evaluation results suggest that 13 solutions suggested by GPT are unsafe and Gemini suggested 9 unsafe solutions as well. We also notice that the error rate varies slightly from 6.6\% (Gemini) to 7.4\% (GPT), suggesting that different LLMs have comparable error rates.

We manually analyzed the problems with the unsafe solutions. Our evaluation results suggest that in most cases (18 out of the 22 cases) the suggested solutions are unsafe because they change the functionality of the involved source code. We call such cases as \textit{semantic bugs} because they change the semantics of the source code.  A typical example is presented in Listing~\ref{listing:semantic}. In this example, GPT conducted an \emph{inline variable} refactoring (Lines 5-7) correctly. It also made some additional changes: Replacing the \textit{if} statements with a ternary operator expression (Lines 10-16). However, the changes as shown in Lines 10-16 are buggy. The expression ``\textit{repo.findRef(startPoint).getName}" may result in a null pointer exception, whereas the original code (and the code refactored by the developer) would not throw a null pointer exception because of the null pointer exception checking on Line 13. 
In total, GPT and Gemini suggested 10 and 6 refactoring solutions respectively that changed the functionality of the involved source code.

\begin{lstlisting}[float, language=diff, numbers=left, xleftmargin=10em, xrightmargin=10em, frame=shadowbox, caption={Suggested Solution That Changes Source Code's Function}, breaklines=true, label={listing:semantic}, belowskip=-1.0 \baselineskip]
    /* From CreateBranchCommand.java in JGit */
    public Ref call() {
        ......
        try (RevWalk revWalk = new RevWalk(repo)) {
 -          %*\colorbox{codered}{Ref refToCheck = repo.findRef(R\_HEADS + name);}*)
 -          boolean exists = %*\colorbox{codered}{refToCheck}*) != null;
 +          boolean exists = %*\colorbox{codegreen}{repo.findRef(R\_HEADS + name)}*) != null;
             ......
 -          String startPointFullName = %*\colorbox{codered}{null;}*)
 -          %*\colorbox{codered}{if (startPoint != null) \{}*)
 -              %*\colorbox{codered}{Ref baseRef = repo.findRef(startPoint);}*)
 -              %*\colorbox{codered}{if (baseRef != null)}*)
 -                  %*\colorbox{codered}{startPointFullName = baseRef.getName();}*)
 -          %*\colorbox{codered}{\}}*)
 +          String startPointFullName = %*\colorbox{codegreen}{startPoint != null ? repo}*)                                                                                     %*\colorbox{codegreen}{.findRef(startPoint).getName() : null;}*)
        ......
    }
\end{lstlisting}

\begin{lstlisting}[float, language=diff, numbers=left, xleftmargin=10em, xrightmargin=10em, frame=shadowbox, caption={Suggested Solution That Introduces Syntax Errors}, breaklines=true, label={listing:syntactic}, belowskip=-1.0 \baselineskip]
    /* From Checker.java in Checkstyle */
    private void processFiles(List<File> files) {
        String fileName = null;
        try {
            fileName = file.getAbsolutePath();
 +         %*\colorbox{codegreen}{final String filePath = file.getPath();}*)  // extracted variable
             ......
        }
        catch (Exception ex) {
             ......
 -          throw new CheckstyleException("......" + %*\colorbox{codered}{file.getPath()}*), ex);
 +          throw new CheckstyleException("......" + %*\colorbox{codegreen}{filePath}*), ex);
        }
        catch (Error error) {
             ......
 -          throw new Error("......" + %*\colorbox{codered}{file.getPath()}*), error);
 +          throw new Error("......" + %*\colorbox{codegreen}{filePath}*), error);
        }
    }
\end{lstlisting}

On the other four cases, the refactoring solutions suggested by LLMs were unsafe because they introduced syntax errors, making the resulting source code syntactically incorrect and thus uncompilable. 
A typical example is presented in Listing~\ref{listing:syntactic}. In this example, GPT conducted an \emph{extract variable} refactoring. However, the refactored code would result in a compilation error because the extracted variable ``\textit{filePath}" (Line 6) is defined inside the \textit{try} block and it cannot be referenced in the \textit{catch} blocks (Lines 12 and 17). The original developer extracted the variable before the \textit{try} block, which did not result in any compilation errors. In total, GPT and Gemini suggested 3 and 1 such kind of unsafe solutions.

Compared against the syntax errors, changes in functionality could be more dangerous because they are often harder to identify. Syntax errors could be found by reliable compilers, and thus they have little chance to influence the final products delivered to end users. However, functional (semantic) changes cannot be identified by compilers. Regression testing has the potential to identify functional changes, but software projects in the wild rarely have sufficient regression unit tests. Consequently, such unintended have the chance to be delivered to end users, leading to serious losses.

\begin{tcolorbox}[
  enhanced,
  rounded corners=all,
  colback=gray!10,
  colframe=gray!100!black,
  colbacktitle=gray!90,
  title=Answer to RQ2-2,
  fonttitle=\bfseries,
  shadow={1mm}{-1mm}{0mm}{black!50!white},
  boxrule=0.5mm,
]
The refactoring solutions suggested by LLMs may be unsafe, i.e., they introduce semantic bugs that change the intended functionality and syntax errors that result in uncompilable code.
\end{tcolorbox}

\section{Detect-And-Reapply Based Bug Mitigation}
\label{section:mitigation}
As suggested by the preceding section, LLM-based software refactoring demonstrates promising results but also suffers from safety problems. The finding highlights the need for careful review and rigorous validation before LLM-refactored code is applied. To this end, in this section, we propose a detect-and-reapply tactic (called \texttt{RefactoringMirror}) to resolve the safety problems.

\subsection{Detect-and-Reapply Tactic}
\begin{figure}[t]
    \centerline{\includegraphics[width=0.72\linewidth]{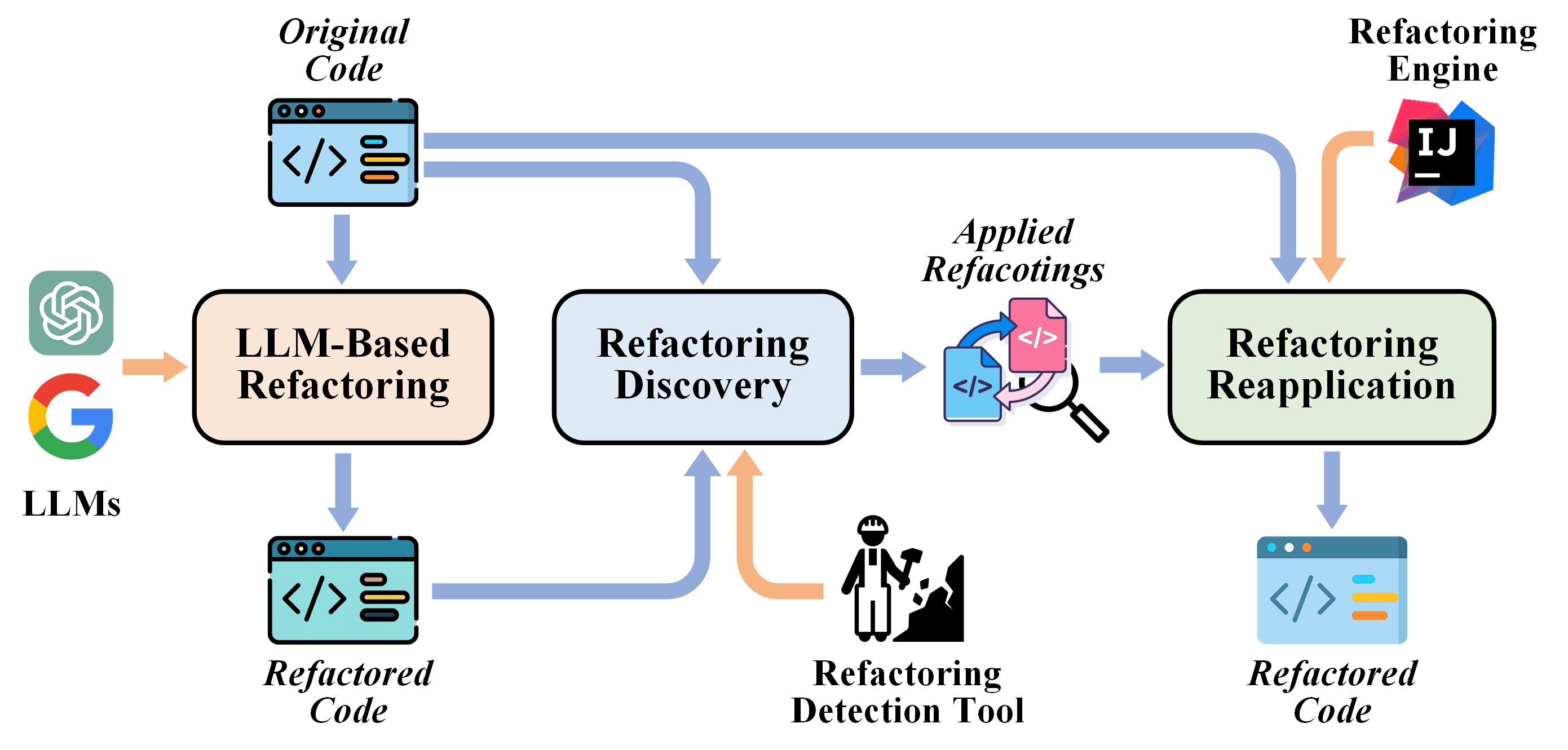}}
    \caption{Overview of RefactoringMirror}
    \label{fig:overview}
\end{figure}

An overview of \texttt{RefactoringMirror} is presented in Fig.~\ref{fig:overview}. As shown in the figure, \texttt{RefactoringMirror} takes two pieces of code as input: The original one (noted as $c$) before refactoring and the improved version (noted as $c'$) generated by a large language model when $c$ is fed into the model. The output of \texttt{RefactoringMirror} is another version of the source code (noted as $\hat{c}$) where all refactorings conducted by the large language model have been conducted whereas unsafe changes conducted by the model are removed.

The key of \texttt{RefactoringMirror} is to identify refactorings conducted by LLMs and to reapply the identified refactorings by well-tested refactoring engines. The workflow is explained as follows.  First, it identifies all refactorings conducted by the large language models with \texttt{ReExtractor}~\cite{lyoubo2024reextractor}, the state-of-the-art refactoring detection tool. \texttt{ReExtractor} takes two versions of the same software system (two documents $c$ and $c'$ in our case) as input, matches code entities across the versions, and infers a list of refactorings according to pre-defined heuristics. Notably, \texttt{ReExtractor} provides only a list of applied refactorings (e.g., an extract local variable refactoring) as well as a list of differences between the two versions. However, it does not present the detailed refactoring solutions (e.g., where the new variable is declared, and which expressions have been replaced by the new variable) that could be fed into refactoring engines to reapply accurately the given refactoring. To this end, for each type of refactoring, we manually design and implement a customized algorithm to extract such detailed refactoring solutions, presented as a sequence of parameters. 

With the extracted detailed refactoring solutions, we request IntelliJ IDEA~\cite{idea2024refactoring} (noted as IntelliJ for short) to reapply the refactorings. IntelliJ was selected because it is one of the widely used refactoring engines, and it is well-known and well-tested. Notably, IntelliJ does not provide simple APIs by which users may automate refactoring with a single API invocation. In contrast, reapplying a refactoring by calling IntelliJ APIs is rather complex, and we have to design and implement a complex algorithm for each type of refactoring.

\subsection{Evaluation}
To evaluate the effectiveness of \texttt{RefactoringMirror}, we applied it to the 22 cases where LLMs introduced bugs. To validate whether \texttt{Refactoring}-\texttt{Mirror} had reapplied all conducted refactorings and to validate whether it avoided all bugs introduced by the large language models, we requested three refactoring experts to manually and independently analyze the results of \texttt{RefactoringMirror}. They should identify all refactoring conducted by \texttt{RefactoringMirror}, compare them against what large language models had done, and validate whether \texttt{RefactoringMirror} had introduced functional changes or syntax errors.

Our evaluation results on the 22 cases suggested that \texttt{RefactoringMirror} was effective in that it successfully reapplied most (94.3\%=33/35) of the conducted refactorings whereas all bugs introduced by the large language models have been avoided. First, we notice that the employed refactoring detection tool \texttt{ReExtractor} was accurate. It successfully identified 33 out of the 35 refactorings (including refactorings explicitly specified by the prompts and additional refactorings automatically conducted by LLMs). \texttt{RefactoringMirror} then reapplied all detected refactorings, and 32 out of the 33 refactorings had been reapplied successfully. Manual checking on the outputs of the \texttt{RefactoringMirror} suggests that \texttt{RefactoringMirror} did not change the functionality of the involved source code or introduce syntax errors, i.e., the modifications are safe. 

\texttt{RefactoringMirror} failed to reapply 3 out of the 35 refactorings because of the following reasons. First, \texttt{ReExtractor} missed 2 refactorings, suggesting that the state-of-the-art refactoring detection tools still deserve substantial improvement. Second, an identified \emph{replace if-else to ternary operator} refactoring has not been reapplied because the employed refactoring engine (i.e., IntelliJ IDEA~\cite{idea2024refactoring}) did not support this type of refactorings. One possible reason is that such refactorings are less popular and thus IntelliJ IDEA does not provide refactoring support for them.

In conclusion, \texttt{RefactoringMirror} can substantially improve the safety of LLM-based refactoring, avoiding all bugs introduced by LLMs. However, because of the limitations of existing refactoring detection tools and refactoring engines, \texttt{RefactoringMirror} may miss some refactorings conducted by LLMs. 
It is acceptable because the proposed approach can effectively alleviate the hallucination issues associated with LLM-based refactoring, ensuring its reliability and preventing potential safety concerns. 

\section{Discussion}
\label{section:discussion}
\subsection{Threats to Validity}
The first threat to external validity is that only a small number of refactoring instances were selected for the empirical study. However, it is challenging to substantially increase the number because it is time and resource-consuming to manually build the ground truth for the selected instances. It took a total of 42 man-days to manually check the number of refactoring opportunities identified by the LLMs and to assess the quality of the refactoring solutions suggested by the LLMs.
To minimize the threat, we employed a reproducible process to select high-quality refactoring instances from 20 open-source projects in different domains.
The second threat to external validity is that we only selected two large language models (i.e., GPT and Gemini) for our study. The conclusions drawn on them may not hold for other large language models. These two models were selected because they had been widely reported as the state of the art~\cite{achiam2023gpt,bang2023multitask,team2023gemini}.
The third threat to external validity is that the empirical study was confined to Java because the state-of-the-art refactoring detection tools did not support programming languages other than Java. 
Although Silva et al. proposed RefDiff~\cite{silva2017refdiff,silva2020refdiff}, a refactoring detection approach that supports multiple programming languages, its detection performance (i.e., precision) and the range of supported refactoring types are significantly inferior to those of the tools selected in this paper~\cite{tsantalis2020refactoringminer}.
Finally, we only investigated within-document refactorings because of the length limitation posed by LLMs on their input and output. The evaluation results reported in this paper may not hold on cross-document refactorings (e.g., \emph{move class} and \emph{move method} refactorings). Notably, the limitation of input length remains a significant challenge for LLMs. As these models continue to advance, they have the potential to handle longer and more complex inputs. Nowadays, one possible solution is to instruct the LLMs to store and remember the context information by feeding them the relevant source code incrementally. After that, we can use the tailored prompt template, along with the source code to be refactored, as input to guide LLMs in generating the corresponding refactored code.

The first threat to construct validity is that the manual construction of the refactoring dataset could be inaccurate. To conduct our study, we requested three refactoring experts to manually validate the refactorings reported by the selected refactoring detection tools, and their validation served as the ground truth for our study. However, manual validation is often subjective, and thus it could be inaccurate. To mitigate this threat, we requested the participants to validate all cases independently and discard cases where inconsistency was found.
The second threat to construct validity is that the manual rating conducted in Section~\ref{section:RQ2-1} could be inaccurate. Note that the assessment of code quality is often challenging and subjective. To minimize the threat, we invited multiple qualified participants to rate refactoring solutions independently. They achieved a Fleiss' kappa coefficient~\cite{fleiss1971measuring} of 0.82, indicating a high level of agreement among them. It may suggest that the resulting ratings could be reliable. 

A threat to statistical validity is the inherent randomness of LLMs’ predictions. Even when the same input and model parameters are used, the results from LLMs may vary due to their stochastic nature. This variability may lead to different outcomes in repeated experiments, making it challenging to achieve consistent results and potentially undermining the reliability of findings. As a result, the reproducibility of experiments involving LLMs may be compromised.
To minimize this threat, we sampled 20 examples from different subject projects for each type of refactoring to reduce potential statistical bias.

\subsection{Limitations}


The first limitation is that the detect-and-reapply tactic may reapply some safe but profitless refactorings. By detecting refactorings conducted by LLMs and reapplying them to the original source code, the tactic guarantees that all changes applied to the original source code are safe, i.e., they may not introduce functional changes or syntax errors. However, not all safe refactorings are beneficial. For example, renaming a method with a nonsense name like ``\textit{amethod}" or ``\textit{m}" is \textit{safe} because it would not result in any bugs. However, it may not help improve the quality of the source code. In contrast, it is harmful in that it could make the renamed method harder to understand. However, neither the proposed detect-and-reapply tactic nor the state-of-the-art refactoring detection tools could distinguish such profitless (or even harmful) refactorings from beneficial ones. 


The second limitation is that although our empirical study in Section~\ref{section:RQ1} and Section~\ref{section:RQ2} confirms that LLMs have the potential to identify refactoring opportunities and to suggest high-quality refactoring solutions, the study does not quantitatively investigate to what extent LLMs can help identify more refactoring opportunities or to what extent LLMs can help reduce the cost of software refactoring. Notably, developers currently conduct refactorings manually or with the help of supporting tools like code smell detection tools (e.g., JDeodorant~\cite{fokaefs2007jdeodorant,tsantalis2011identification} and PMD~\cite{pmd2024pmd}), refactoring solution advisors (e.g., RefBERT~\cite{liu2023refbert} and EM-Assist~\cite{pomian2024together}), and refactoring engines (e.g., IntelliJ IDEA). In this paper, we do not empirically compare LLM-based refactoring against such manual or semi-automated approaches. Consequently,  it remains unclear to what extent developers can benefit from LLM-based refactoring concerning the effect (i.e., quality improvement caused by conducted refactorings) and the cost (e.g., human cost and computational expense) compared against the state of the art.

Although specifying expected refactoring types may increase LLMs' success rate in identifying refactoring opportunities, developers may not know exactly the expected refactoring types in advance. In this case, the developers should either try each of the popular refactoring types one by one or add all such refactoring types as expected refactorings. 
As introduced in Section~\ref{sub:RQ1-4}, explicitly specifying the refactoring subcategories and narrowing the search space may further improve LLMs' success rate in identifying refactoring opportunities. However, developers may not know exactly the expected refactoring subcategories in advance. Therefore, we suggest that developers should try each of the proposed refactoring subcategories one by one to determine whether the given source code contains such refactoring opportunities. Moreover, the proposed refactoring subcategories in this paper may not be exhausted. Consequently, we suggest that developers should customize dedicated prompts for the refactoring subcategories they are interested in, based on the prompt templates provided in this paper.

One limitation of the \texttt{RefactoringMirror} is the dependency on the capabilities of existing refactoring engines, which do not support a comprehensive range of refactoring types. Therefore, they may not accommodate the diverse and potentially novel refactorings suggested by LLMs. This discrepancy can lead to situations where the \texttt{RefactoringMirror} identifies valid refactorings that cannot be reapplied automatically due to the constraints of the current refactoring engines. To mitigate this limitation, we propose an interactive tactic where both the outputs of \texttt{RefactoringMirror} and the modifications made by LLMs are presented to users. This tactic allows users to make informed decisions on whether to accept, modify, or reject the suggested changes. By involving the user in the decision-making process, we leverage human expertise to bridge the gap between the advanced refactoring capabilities of LLMs and the current technical constraints of refactoring engines.

\subsection{Implications}
LLMs are best used as suggestive auxiliary tools rather than precise and reliable code refactoring tools. On one side, LLMs can identify many refactoring opportunities and suggest many high-quality refactoring solutions, and thus they could be used as smart assistants. On the other side, LLMs often miss refactoring opportunities, suggest nonsense solutions, and conduct refactorings incorrectly (resulting in bugs), which makes the current form of LLMs unsuitable as a reliable refactoring engine. 

The study's findings demonstrate that LLMs possess the potential to autonomously identify refactoring opportunities. However, without clear and concrete guidance, LLMs may 
struggle to accurately pinpoint such opportunities. Therefore, we suggest that developers avoid relying on generic prompts when working with LLMs for refactoring tasks. Instead, they could systematically 
apply the tailored prompt templates we provide for each type of refactoring, as outlined in Section~\ref{sub:RQ1-4}. By applying these prompts one by one, LLMs 
can be effectively guided to identify potential refactoring opportunities in a sequential and structured manner.
Once a refactoring opportunity is identified, LLMs can generate corresponding refactoring suggestions, which developers can review and decide whether to implement the suggested changes. 
Notably, LLMs serve as only one component of the automated refactoring process. We believe that as LLM capabilities continue to advance, 
the proposed prompt templates and stepwise strategies will increasingly align with developers’ refactoring objectives, facilitating a more efficient and targeted refactoring workflow.

Post-processing is critical for LLM-based software engineering tasks. Large language models are suffering from the well-known hallucination problem~\cite{xu2024hallucination,tonmoy2024comprehensive,bang2023multitask}, making it difficult to ensure their reliability. The problem is especially serious when LLMs are employed to modify source code where a minor mistake may result in serious bugs in the resulting software systems. In this case, a well-designed post-processing could identify or even exclude such minor mistakes, significantly improving the reliability of LLM-based software engineering tasks. 
The approach, \texttt{RefactoringMirror}, proposed in this paper serve as a good example of how post-processing can be integrated into LLM-driven refactorings. Future work can further develop and expand the proposed approach to enhance the reliability of LLMs.

\section{Related Work}
\label{section:relatedwork}
\subsection{Deep Learning-based Refactorings}
With the advances in deep learning technologies, many researchers have investigated deep learning-based refactoring. Liu et al.~\cite{liu2019deep} were the first to apply deep learning techniques to software refactoring. Their major contribution is a novel approach to synthesizing large-scale training data for neural networks that are designed to detect code smells and suggest refactoring solutions. However, deep neural networks trained with synthetic data may only learn to identify these artificial smells, rather than real-world code smells. To this end, Liu et al.~\cite{liu2023deep} proposed to automatically collect code smells and refactorings from real-world projects, which could be directly employed as high-quality training data. Barbez et al.~\cite{barbez2019deep} were the first to exploit historical and structural information to identify god class smells with the help of deep learning techniques. Kurbatova et al.~\cite{kurbatova2020recommendation} proposed a hybrid approach to identify feature envy smells that combines deep learning techniques (i.e., code2vec~\cite{alon2019code2vec}) and an SVM-based classifier. Similarly, Cui et al.~\cite{cui2022rmove} applied code2vec (and code2seq~\cite{alon2019code2seq}) to transform methods into numeric vectors, and utilized graph embedding techniques to represent the dependencies between the method and other methods. With the resulting embeddings, they leveraged traditional machine learning techniques (like Naive Bayes) to suggest move method refactorings.

Tufano et al.~\cite{tufano2019learning} conducted the first empirical study on the capability of neural machine translation (NMT) models to learn and apply code changes (e.g., refactorings) made by developers. Their evaluation results suggest that NMT models can replicate up to 36\% of the changes made by developers. Liu et al.~\cite{liu2019learning} proposed an automated approach based on deep learning and information retrieval to identify methods whose names are inconsistent with the corresponding method bodies. Desai et al.~\cite{desai2021graph} proposed to automatically refactor a monolith application into multiple microservices using deep learning techniques. Liu et al.~\cite{liu2023refbert} proposed a two-stage pre-trained framework based on BERT architecture~\cite{kenton2019bert} to suggest appropriate names for poorly named variables. Vitale et al.~\cite{vitale2023using} proposed an approach to automatically improve code readability by fine-tuning the T5 model~\cite{raffel2020exploring} with readability-improved commits mined from software repositories.

\subsection{LLM-based Software Engineering}
With the emergence of large language models (LLMs), LLMs have been applied to multiple software engineering tasks, demonstrating their remarkable capabilities. Prompts serve as structured inputs that guide LLMs to generate more relevant and context-specific outputs. White et al.~\cite{white2023prompt,white2024chatgpt} explored the design of prompt patterns to enhance the interaction between developers and LLMs. Their work identified and classified different prompt patterns that could improve the performance of LLMs in software engineering tasks, making these models more accessible and efficient for developers. Dong et al.~\cite{dong2024self} proposed a self-collaborative framework for code generation tasks based on ChatGPT~\cite{chatgpt2024blog}. The evaluation results suggest that their framework outperforms the state-of-the-art code generation benchmarks and even GPT-4. Liu et al.~\cite{liu2023your} proposed a program synthesis evaluation framework for evaluating the correctness of LLM-generated code. Sch{\"a}fer et al.~\cite{schafer2023empirical} exploited LLMs to generate unit tests automatically, achieving a median statement coverage of 70.2\% and branch coverage of 52.8\%, outperforming the state-of-the-art feedback-directed test generation techniques. Xia et al.~\cite{xia2023keep} proposed a conversation-driven automated program repair approach based on ChatGPT. The approach successfully fixed 162 out of 337 bugs on the Defects4J dataset~\cite{just2014defects4j}.

Liu et al.~\cite{liu2023improving} investigated the capability of ChatGPT in code generation tasks. Their findings reveal that the carefully guided ChatGPT outperforms other fine-tuned pre-trained language models (e.g. CodeT5~\cite{wang2021codet5} and CodeGPT~\cite{lu2021codexglue}). 
Guo et al.~\cite{guo2023exploring} explored the potential of ChatGPT in automated code refinement tasks, and the evaluation results suggest that ChatGPT outperforms the state-of-the-art code review tool. Ouyang et al.~\cite{ouyang2023llm} investigated the non-determinism of ChatGPT in code generation tasks. Their evaluation results suggest that setting the temperature to zero tends to generate more deterministic code, which is consistent with the observations made by Guo et al.~\cite{guo2023exploring}. Sun et al.~\cite{sun2023automatic} assessed the performance of ChatGPT in code summarization tasks, highlighting its significant inferiority to the state-of-the-art models in this task. Wu et al.~\cite{wu2023large} investigated the capability of GPT-4 in fault localization tasks. Their evaluation results on the widely-used Defects4J dataset suggest that GPT-4 outperforms the existing approaches and GPT-3.5. Noever et al.~\cite{noever2023can} evaluated the capability of GPT-4 in detecting software vulnerabilities, demonstrating that GPT-4 identified approximately four times as many vulnerabilities as its competitors and also provided viable fixes for each one. Vaithilingam et al.~\cite{vaithilingam2022expectation} presented a user study on the usability of Copilot~\cite{github2023copilot}, an LLM-based programming assistant. The study found that most participants preferred to use it because it provided a useful starting point and saved the effort in online search.

\subsection{LLM-based Software Refactoring}
Given the remarkable proficiency of LLMs in various software engineering tasks, several studies~\cite{chouchen2024so,tufano2024unveiling,alomar2024refactor,deo2024analyzing} have investigated how developers exploit LLMs for software refactoring. These studies emphasize the role of LLMs in improving software quality through refactoring and detail how developers interact with LLMs to specify and execute refactoring. 
Complex programs often present challenges in terms of readability and maintainability. To address these issues, Shirafuji et al.~\cite{shirafuji2023refactoring} proposed an approach that uses few-shot examples to prompt LLMs to suggest simplified versions of such programs. Their evaluation results suggest that the simplified programs significantly reduce cyclomatic complexity and the number of lines in code while preserving semantic correctness.
Existing Transformation by Example (TBE) techniques are employed to automate code changes. However, their effectiveness may struggle with unseen code change patterns. To this end, Dilhara et al.~\cite{dilhara2024unprecedented} proposed PyCraft, which utilized LLMs to generate semantically equivalent and yet previously unseen variants of change patterns. With such variants, PyCraft could significantly increase the effectiveness of TBE techniques and expedite the software development process. Pomian et al.~\cite{pomian2024together,pomian2024assist} proposed an LLM-based approach, called \texttt{EM-Assist}, to suggest safety \emph{extract method} refactorings for decomposing long methods. As the initial step, LLMs were repeatedly requested to generate \emph{extract method} refactoring solutions. To ensure the suggested solutions were valid, i.e., the refactored code could compile successfully, the authors utilized IDE's API to validate whether the suggestions meet the refactoring preconditions for the extract method. Subsequently, the program slicing techniques were employed to enhance the LLMs' suggestions by adjusting the code fragments. Although \texttt{EM-Assist} received favorable feedback from many industrial developers, its scope is limited to a specific type of refactoring and does not conduct a comprehensive evaluation of LLMs' refactoring capabilities. Depalma et al.~\cite{depalma2024exploring} explored the refactoring capabilities of ChatGPT, revealing that it can effectively refactor code and offer insights into the refactored code. However, their study lacked a quantitative evaluation of how well LLMs perform in automate software refactoring compared to human experts. To this end, we construct a high-quality dataset consisting of refactorings conducted by developers in real-world scenarios, and evaluate for the first time whether LLMs have the potential to identify refactoring opportunities accurately. Furthermore, we propose a detect-and-reapply tactic to mitigate the hallucination issues associated with LLMs, ensuring the reliability of LLM-based refactoring.

Existing studies primarily focus on using LLMs to suggest and execute specific refactorings automatically, yet there is a lack of comprehensive empirical study on the refactoring capability of LLMs compared to those of human experts. Moreover, we propose the detect-and-reapply tactic to mitigate issues with LLM-generated \emph{hallucinations}. Notably, in this paper, the hallucinations refer to instances where LLM-driven refactorings result in syntax errors or change the intended behavior of the code. This definition differs from that of Pomian et al.~\cite{pomian2024together,pomian2024assist}, who developed EM-Assist to decompose long methods and defined LLMs' hallucinations as refactoring suggestions that fail to meet IDE's refactoring preconditions and refactoring solutions that involve only one single-line statement or the whole method body. Our paper aims to complement these studies by conducting a rigorous empirical study on LLM-based software refactoring. 

\section{Conclusions and Future Work}
\label{section:conclusion}
In this paper, we conduct an empirical study to investigate the potential of LLM-based software refactoring. Our findings suggest that with proper prompts, LLMs have the potential to identify 86.1\% of the refactoring opportunities with high accuracy. LLMs also have the capability to suggest high-quality refactoring solutions for the identified refactoring opportunities, and such solutions are comparable to those constructed manually by human experts. However, our evaluation results also suggest that LLMs may introduce some dangerous changes that either reduce the performance of the source code or introduce function and/or syntax errors. To this end, in this paper, we propose a detect-and-reapply tactic to void buggy changes. Our evaluation results suggest that it successfully avoided all bugs initially introduced by the evaluated LLMs. 

Our empirical study demonstrates the potential of LLM-based refactoring, which may inspire further research in this line. It could be interesting in the future to investigate how to assess the refactoring opportunities and refactoring solutions suggested by LLMs, and thus we can pick up only beneficent refactoring opportunities and high-quality refactoring solutions. It is also potentially fruitful to investigate how to improve LLM-based refactoring by pre-processing and post-processing. For example, according to the few-shot learning techniques, it could be helpful to retrieve examples from a corpus according to the to-be-refactored source code and append them to the prompt. Our evaluation results may also encourage the integration of LLM-based refactoring into code review. The LLMs' suggestions could be fed to developers, and they would manually validate such suggestions and conduct refactorings, which helps avoid unsafe refactorings.


\bibliographystyle{ACM-Reference-Format}
\bibliography{sample-base}

\end{document}